# Going Negative Online?
# --A Study of Negative Advertising on Social Media

**MY498-Capstone Project**

**Masterthesis**

**Targeted academic degree:**
**Master of Science (M.Sc)**
**in Applied Social Data Science**

**Department of Methodology**
**London School of Economics and Political Science**

**Candidate Number: 8633**
**Supervisor: Pablo Barberá, Blake Miller**
**Word count: 8615**
**Date: 08.08.2019**

**Abstract:**

In the digital age, political advertisers are increasingly choosing to advertise on social media, as social media ads can provide a more advanced micro-target tool for targeting individuals and provide external landing pages that allow advertisers to better engage their audiences. Negative advertising has also become less obvious due to the ad confidentiality policies on social media. A growing number of empirical studies suggest that negative advertising is effective in campaigning, while the mechanisms are rarely mentioned. With the scandal of Cambridge Analytica and Russia's intervention behind Brexit and Trump’s election, people have become aware of the political ads on social media and have pressured congress to restrict political advertising on social media. Following the related legislation, social media companies began disclosing their political ads archive for transparency during the summer of 2018 when the midterm election campaign was just beginning. This research collects the data of the related political ads in the context of the U.S. midterm elections since August to study the overall pattern of political ads on social media and uses sets of machine learning methods to conduct sentiment analysis on these ads to classify the negative ads. A novel approach is applied that uses AI image recognition to study the image data. Through data visualization, this research shows that negative advertising is still the minority; Republican advertisers and third party organizations are more likely to engage in negative advertising than their counterparts. Based on ordinal regressions, this study finds that anger-evoked information-seeking is one of the main mechanisms causing negative ads to be more engaging and effective rather than the negative bias theory. Overall, this study provides a unique understanding of political advertising on social media by applying innovative data science methods. Further studies can extend the findings, methods, and datasets in this study, and several suggestions are given for future research.

## 1. Background

With the development of social media and information technology, political advertising on social media tends to be more targeted and more efficient. Social media has grown to become one of the main battlefields for candidates. In the 2016 US presidential elections, the two candidates spent approximately 81 million dollars on their promotions on Facebook according to Colin Stretch, Facebook's general counsel. “Going negative” is the main impression that many US citizens have of this election, as both of the candidates and their supporting media used many negative messages in this election. Some argue that the pro-Trump fake messages on social media are far more in terms of amount and negativity than those of Hillary, which may have contributed to Trump’s victory (Allcott & Gentzkow, 2017). Many scholars argue that negative sentiments and controversial words in political messages can easily draw attention, which causes the candidate to receive more likes, comments, and shares (Alashri et al., 2018), making his or her words spread much further than his opponents and is even boosted by the ad distribution model of these social media sites when the message is in the form of an ad. Meanwhile, the scandal of Russian agents’ involvement in political advertising online and Cambridge Analytica’s misuse of users’ data on social media shocked the whole democratic world. However, political advertising on social media remained a black box without any open data for a long time. People began to realize the importance of the transparency of political advertising online after these controversies and scandals. Recently, to better restrict and regulate the political advertising on social media, lawmakers in United States introduced the Honest Ads Act to address with those concerns about political ads on social media. In response to this bill and the scandals, social media companies, such as Facebook, Google and Twitter, disclosed their dataset for political ads, which provides the data for this research.

### 1.1 Definition of social media and ad rank

The definition of social media in this paper refers to a wide range of online platforms that enable people to access information, communicate and interact with each other, including Facebook, WhatsApp, Twitter, Google and so on (Mangold & Faulds, 2009). Social media sites have grown to become primary information sources for people in most countries and tend to become battlefields during elections.

The most important mechanisms that make social media different from the traditional media are the Real-time bidding (RTB) design for advertisers and the engagement-based model that social media sites use for ad ranking and distributing. Social media sites use a complicated algorithm to optimize the display of ads by adding weight to the engagement performance of ads (e.g., clicks) rather than just relying on the payment amount from advertisers. As a result, if the content of an ad is more engaging, this ad will have a better ranking and more purchasing power, making it acquire more impressions with a lower price. The metric of the engagement performance of ads is called the eCPM (effective cost per mile), which is introduced in the following data collection section, and is the measure of audience engagement used in this research. Additional details about the micro-targeting tools and ad ranking on social media are shown in Appendix (1).

## 2. Literature Review and Research Questions

### 2.1 General patterns of political advertising online:

Kaid & Postelnicu (2005) first studied the political ads on the Internet and compared them with the traditional ads and found that their effect is different from that of television ads. For the negative political advertising online, Druckman, Kifer & Parkin (2010) conduct a comprehensive analysis of the negative online advertising and find that the likelihood of candidates' campaigns going negative online is similar to that of going negative on TV. Ridout, Franklin & Branstetter (2010) examine the political ads during the presidential election on Youtube and find that the most viewed ads, by both traditional and non-traditional actors, are in various formats. Recently, Edelson, Sakhuja, Dey & McCoy (2019) studied the disclosed political ads archives from Facebook, Google and Twitter and collected ad content data to describe the general pattern of online political advertising in the United States. However, only the basic patterns in terms of spending, ad types, and distribution are reported. Negativity, sentiments and other distribution patterns are not covered.

### 2.2 Effect of Political Advertising:

The most fundamental question regarding political advertising is whether political advertising can affect voters' behaviour. The literature on this topic gives mixed results, in terms of both theories and empirical studies. On the one hand, many scholars (i.e., Bennett & Iyengar, 2008) support the minimal effects theory that originated in 1940s, which suggests that political campaigns only have minimal effects on voters, as voters already had clear positions on politics and issues. On the other hand, recently, a growing literature endorses the hypodermic needle theory, or the magic bullet theory, which suggests that an intended message will be accepted by the targeted receiver and affect the receiver's behaviour. For example, Atkin & Heald (1976) empirically test the effect of political advertising by collecting data from random interviews and the findings suggest that higher exposure to political ads is correlated to higher agenda priorities regarding issues and candidate attributes, and the effect on voter behaviour was mildly associated with political advertising frequency. However, this type of empirical research on the effect of political advertising is considered highly endogenous by many studies (i.e., Gordon et al., 2012), since there are many unobserved variables, such as the fundraising, decision-making processes of candidates and parties. Some methods to resolve these issues are found in later empirical research, for instance, Gordon & Hartmann (2013), Chung & Zhang (2016) use the last year's advertising prices as instrumental variables to study the causal relationship between voter behaviour and political advertising and point out that political advertising does affect the voters' preferences for candidates. Wang, Lewis, and Schweidel (2018) use a border strategy to resolve the endogeneity concern and find that advertising affects audience behaviour.

### 2.3 Effect of Negative political advertising:

Regarding the effect of negative political advertising, conventional wisdom suggests that negative advertising is an effective strategy in political campaigns so that it is increasingly widespread, while the empirical results are mixed. On the one hand, some empirical studies on the effects of negative advertising (i.e., Lau et al. 1999, 2007) find that there is not strong evidence that negative advertising is a more effective way to change voter behaviour. On the other hand, many studies find the effect of negative advertising on voter behaviour is strong, at least under some conditions. Newhagen & Reeves (1989) compare the effects of positive ads and negative ads, and find that negative ads are more effective in improving recognition memory. Garramone, Atkin, Pinkleton and Cole (1990) use an experiment to compare the effects of negative ads and positive ads, and the findings suggest that negative ads result in stronger discrimination of the candidate's image and greater political polarization compared with their positive counterparts. Johnson-Cartee and Copeland (1991) suggest that the negative ads strategy has a negative effect on the opponent but also the sponsor. Ansolabeher and Iyengar's study (1997) suggests that exposure to negative advertisement would reduce the voter turnout, while exposure to negative ads can strengthen potential partisan sentiments. Ansolabehere, Iyengar and Simon (1999) combine experimental and aggregate survey data, and find that negative advertising can demobilize voters. Later, with the arrival of the digital era of political advertising, empirical studies increasingly found that negative advertising is more effective. For instance, Ceron and Adda (2016) studied the 2013 Italian general election and found that a negative digital campaign has positive effects on voters' voting intentions and this effect is strengthened when the attacker is also under attack by his or her rivals. Wang, Lewis & Schweidel (2018) use a border strategy analysis to study the effect of political campaigns on social media in US senatorial elections given the advertising discontinuities along DMA borders provide a natural experiment, and found that the negative advertising sponsored by candidates has a positive effect on vote share, while the ads sponsored by PACs (Political Action Committees) are significantly less effective.

**2.4 Mechanisms behind negative political advertising:**

One commonly believed mechanism of how the negative advertising works is explained by the negative bias theory. The psychology studies (Carretié, Mercado, Tapia & Hinojosa, 2001; Rozin & Royzman, 2001) find that negative events cause faster and more prominent responses than non-negative events for people, as people give greater weight to negative things. Negative advertising takes advantage of this negative bias to change people's minds and then their behaviours, for instance, there is substantial evidence that politicians can involve highly emotional content in their ads with the intention of evoking interest and then politically relevant behaviours, such as partisan support and voting.

Another mechanism that can explain the effect of negative advertising is that some particular emotions in the negative ads are engaging and evoke political related behaviours rather than overall negativity. For instance, anxiety was found to be very influential in changing the political behaviours of audiences, making them search for more political information, remember more political information, be more likely to join political events and more likely to vote (e.g., Brader 2006; Lerner et al. 2003; Marcus, Neuman, and MacKuen 2000; Redlawsk, Civettini, and Lau 2007). Valentino and Brader (2011) found a distinctive influence of anger in political mobilization by a randomized experiment. However, Ryan (2012) first pointed out a possible explaining mechanism of why negative messages online work from an emotion, and he finds that anger can efficiently increase the probability of people's information seeking and then increase their tendency to click a political website that increases psychological incentives for political communication and other political related behaviours.

**2.5 Summary of Literature Review:**

In this research, I assume that the negative political advertising is effective in affecting voters' behaviour, as many previous empirical studies have shown that political advertising is effective and negative advertising is even more effective in affecting voter behaviours, such as increasing the vote share for a candidate and decreasing voter turnout of the competitor. The reason that negative advertising is effective in affecting voter behaviours is that negative ads are more engaging and can stimulate information seeking due to some mechanisms. The negative bias theory and the theory that specific emotions evoke information-seeking are the two possible mechanisms of explaining the improvement of engagement through negative advertising. Therefore, I investigate the overall pattern of political advertising online and explore how the advertising influences the voter response from a sentiment analysis perspective, testing the two possible channels in the online advertising environment with the assumption that negative advertising matters. In summary, this research intends to study the negative campaigns online in the elections and find the mechanism of how negative political ads on social media affect voter behaviour.

**2.6 Research Questions:**

Based on the related literature above, I intend to further explore political advertising on social media in this study. I have two sets of research questions involving (1) descriptive questions and (2) explanatory questions. The first set of questions includes the following:

1. how frequently do political ads go negative on social media?
2. which types of political advertisers are more likely to use negative ads?
3. what are other patterns or key differences across campaigns on social media?
4. what content in the ads are more likely to go negative?

The explanatory question is as follows: how do negative ads affect voter behaviour? If the negative bias theory is correct, more negativity in an ad should result in a higher likelihood of responses to it; therefore, I can obtain my first hypothesis. If the theory that anger evokes information-seeking is correct, an ad that evokes more anger should result in a higher likelihood of responses to it; therefore, my second hypothesis is shown below.

Hypothesis 1: Going negative increases the likelihood of responses in political advertising.
Hypothesis 2: The anger that is evoked from negative ads increases the likelihood of responses in political advertising

**3. Data Collection**

I select political ads from Google for this research, because Google's market share is the largest in the American online advertising market, occupying 37.2% of the total market share (eMarketer, 2018) and even more of the online political ad market share. Another reason is that Google only allows political advertising related to federal elections, and the other political ads, such as issues ads, are not included, which is more suitable for this study. With the big data analytical tools, all of the electoral ads on Google during the campaign of the 2018 United States midterm elections (from 1st August to 6th November) are collected for my analysis, which includes data of 36,814 ads. This should be sufficient to draw the overall pattern of only the political advertising and test the effect of negativity and emotions on engagement.

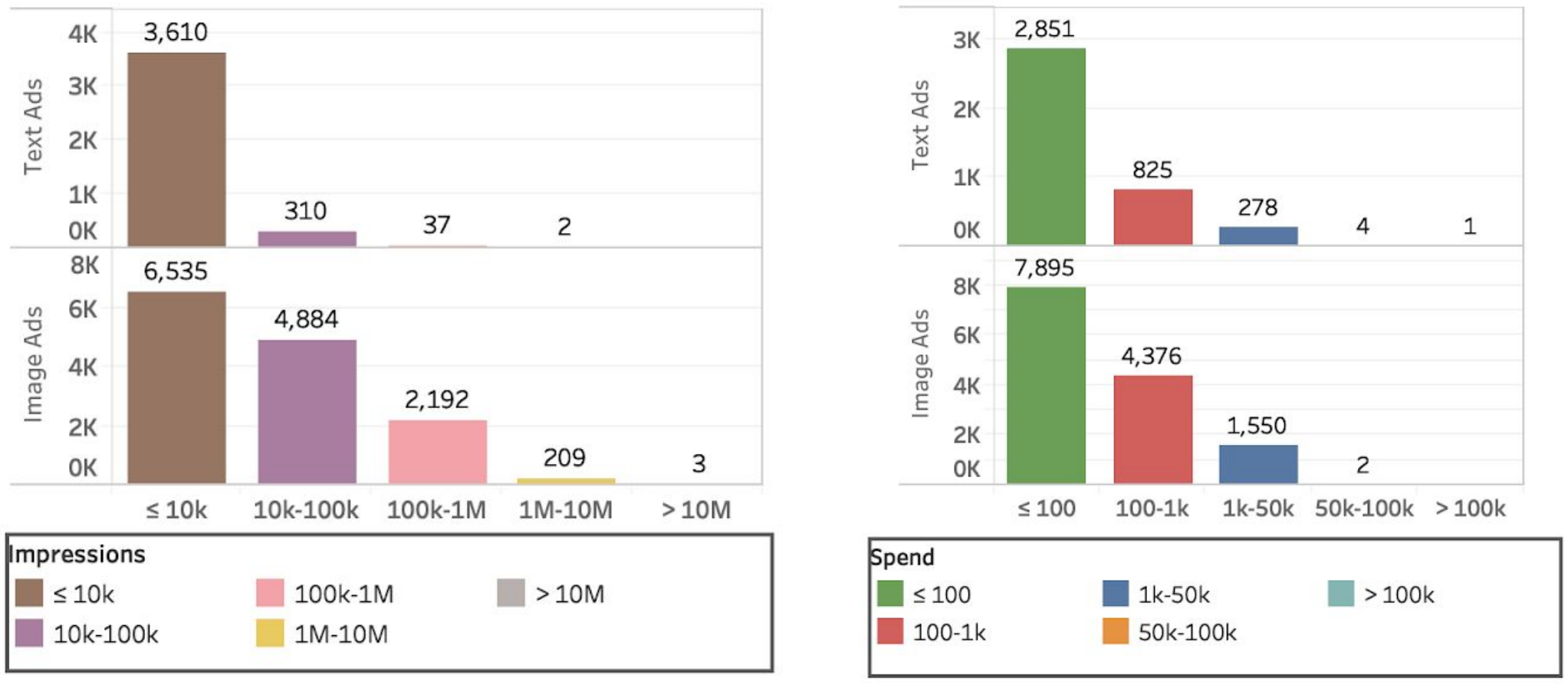

Figure 1: Distribution of ads by size

### 3.1 Ad statistics:

Google directly released their ad archive recently in a Big-Query format on Google Cloud for the public. However, the ad content data were not displayed in the dataset, but were given through a link to a summary page for each ad. As a result, I extract all of these ad summary pages by web-scraping to collect the detailed text or image content for this research. However, due to a technical limitation, I was unable to analyse the video ads efficiently. Thus, I only collect text and image ads from Google; and the total size of the ads in this subset is 20,769 rows. Google's ad data also include the time period, total spending and impressions for each ad. It provides the targeted demographic data of the audience, including the age group, gender and location. The total impression (people reached) is also shown by range instead of giving an exact number. Specifically, the total impression are presented in the following 5 ranges: <10k, 10k-100k, 100k-1M, 1M-10M, >10M. The 5 ranges of spending are displayed as follows: < 100, 100-1k, 1k-50k, 50k-100k, >100k. The distributions of ads by the impressions and spending are shown in Figure 1. It is clear that the two distributions of ads do not match each other perfectly, though these two distributions are quite similar and both suggest a decreasing distribution. Based on this difference, I find that some ads outperform the others that have lower spending but higher impressions and vice versa, and this variation in ad performance can measure the engagement of the ads.

### 3.2 eCPM:

This variation is called the eCPM (technically, "effective cost per mille"), which means the effective cost per thousand impressions, which is a metric of expenditure for digital ads and a measure of the performance of an ad in practice. With the impression and spending data, I estimate a rough eCPM ratio for each ad. The original formula is eCPM = S/I, where "I" refers to total impressions in thousand and "S" is total spending. In this study, I estimate the rough eCPM level by simply dividing the level of impression by the level of spending. In the following analysis, I use the estimated eCPM as a key measure of engagement of an ad, since the engagement performance (click-through rate) determines the eCPM (Chapelle, 2014; McAfee 2011). Additionally, according to Google's explanation of its undisclosed ad rank algorithm[1], ad quality is the most important factor in the ad rank, given a similar bid amount

1 detail: https://support.google.com/google-ads/answer/1752122

and bid time, while the eCPM is the only available measure of ad performance. Thus, I assume that the bid amounts of all advertisers on Google are competitive and are above the minimum ad rank threshold during the election campaign period and attribute the variance of eCPM to the difference in the engagement performance in the ad content. The estimated eCPM is an ordered discrete variable in nature that can be applied to my ordinal logistic regression model in the next part.

### 3.3 Advertiser attributes:

In addition, I manually categorize all the advertisers by organization partisan (Democratic, Republican or Independent/unknown), election state, and type. The advertiser type includes political candidates and third-party organizations that includes the Political Action Committee (PAC), super PAC, Union, Party Committee, Agency, etc., for this dataset. I label each advertiser according to the source from OpenSecrets.org and the advertiser's website. If an advertiser has no clear partisan affiliation or usually supports candidates from both parties that work for the same interest, it is also labelled as 'Independent/unknown'.

### 3.4 Ad Content:

The ad content for each ad is accessible through a given URL link, and the ad content sample pages can be found in appendices 2 & 3. Thus, I collect these content data from the ad content link through web-scraping. This raw data will be used to conduct exploratory and sentiment analysis, which will be introduced in the data analysis section.

This ad dataset includes 6491 rows of data of text ads that ran from August 1st to the election day, November 6th, on Google. After scraping the content data, only 61% of the contents of the textual ads are collected. This result is very similar to other researchers' scraping results (i.e., Edelson, Sakhuja, Dey & McCoy, 2019). The other 39% of the textual ad content is missing mainly due to blockage by Google, which displayed the message "*Policy violation This ad violated Google's Advertising Policy*." These blocked ads are very likely to include negative ads, since Google's advertising policy strictly regulates the word use in the ad content[2]. Thus, there are 3959 documents in my corpus, and each document is a

2 Details about Google's advertising policy: https://support.google.com/adspolicy/answer/6008942?hl=en

textual ad displayed for target audiences on Google. The word cloud below shows the most frequent words in this corpus after removing stop words.

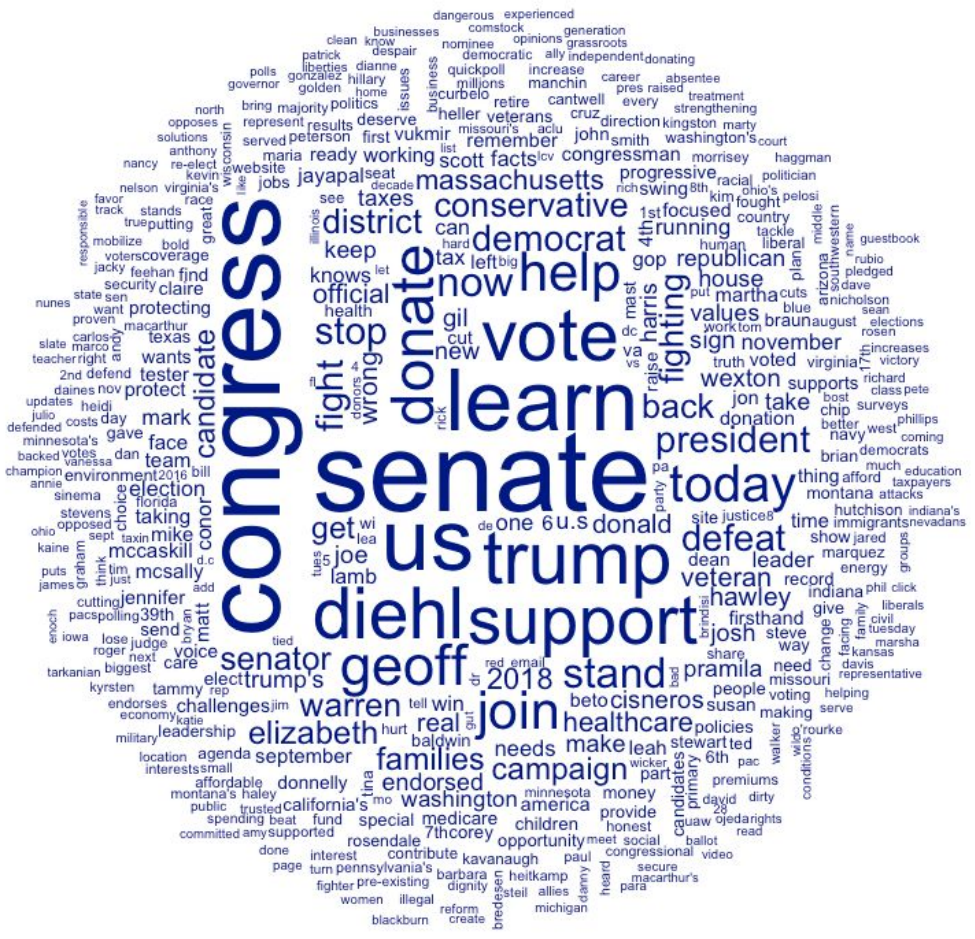

Figure 2: Word Cloud of Text Ads

The word cloud graph shows that, as expected, the political ads on social media during this period are mainly about the midterm elections, as “senate” and “congress” are clearly the most frequently mentioned words. It also indicates some main intentions or topics of the ads, such as “donate”, “vote”, “learn”, “support”, and “Trump”.

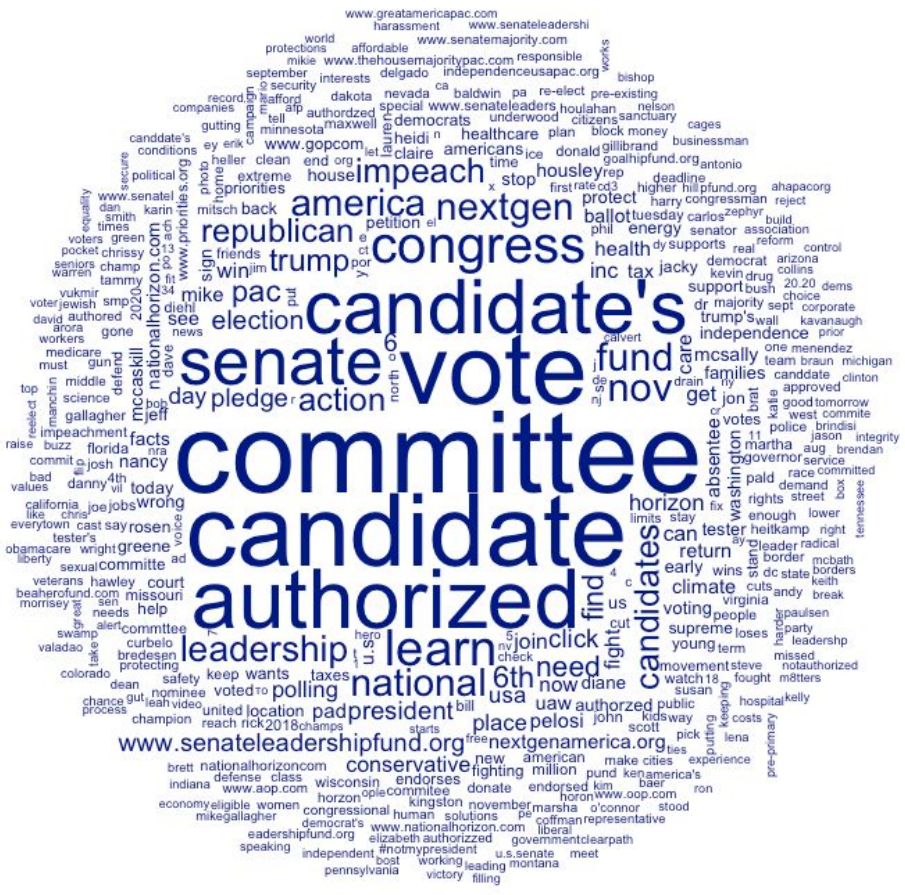

Figure 3: Word Cloud of Text in Image Ads

Regarding the image content, the scraper successfully collected 94.5% of the images in the image dataset. The other minor part, i.e., 5.5%, of the textual images are missing because of the following two main reasons other than blockage: 1. invalid pages; 2. some video ads are misplaced in the image dataset. As a result, I collect relatively comprehensive image content. However, I would like to suggest that Google should clean this dataset. The text content in the image ads is also detected and shown above in Figure 3. The most frequent words are very similar, however, the word “authorized”, which is mentioned frequently, is not covered in the text ads. This is because almost all the third party advertisers included the following sentence in the image ads, as the new law requires: “The ad is not authorized by any candidate”.

## 4. Data Analysis Methodology

### 4.1 Sentiment Analysis

To classify the negative political ads and measure the negativity, I apply a sentiment analysis method in this research, since negative political ads usually evoke negative emotions in their content. The negativity of an ad is measured by the negativity score of its content (both for the text and image content). With the negativity and emotions of both the text and image ads detected and scaled, I can conduct consistent descriptive analysis about the ad content and compare the results by data visualization. Additionally, the overall effect of negative advertising on users’ engagement (measured by eCPM) through negativity bias or specific emotions is tested with regression analysis.

Specifically, to classify and scale the sentiment and emotions in the text ad, I apply an NRC Word-Emotion Association lexicon (Appendix 4), which codes 2 sentiments, i.e., negative and positive, and 8 specific emotions, i.e., anger, anticipation, disgust, fear, joy, sadness, surprise, and trust. According to this lexicon, the real-valued score that measures the degree of the eight emotions evoked and the corresponding sentiment in a document are calculated[3]; thus, I can test the negativity and 8 emotions in the following text.

Based on the annotation values of emotions and sentiment, I establish the following ordinal logistic regression models to test whether negativity can affect audience engagement (measured by eCPM), and which specific emotions work in affecting eCPM.

$$eCPM = \beta_0 + \beta_1 negativity + \beta_2 t + \beta_3 party + \beta_4 thirdparty + \mu$$

$$eCPM = \beta_0 + \beta_1 anger + \beta_2 joy + \beta_3 sadness + \beta_4 fear + \beta_5 surprise + \beta_6 anticipate + \beta_7 trust + \beta_8 disgust + \beta_9 t + \beta_{10} party + \beta_{11} thirdparty + \mu$$

3 Details about the dictionary and scores of emotions: http://saifmohammad.com/WebPages/lexicons.html

For the image data, I use a trained machine learning model on Google cloud to conduct image recognition that extracts useful data from each ad image for analysis. The Google Cloud Vision provides an advanced trained image recognition model to detect the labels (topics, see appendix 5), sentiments, and text through API connection[4]. However, this model is not open source; therefore, the detailed machine learning algorithm cannot be checked. Note that Google Cloud Vision annotates the likelihood of emotions in each ad as the following 5 levels: "very unlikely", "unlikely", "possible", "likely", and "very likely". However, it only detects 4 emotions, i.e., anger, sorrow, joy, surprise as shown in appendix 6. As a result, I calculate the negativity score based on the ordinal likelihood of emotions. The negativity score is calculated to scale the sentiment in two different equations as shown below:

$$negativity = anger + sorrow - joy \quad \text{(equation 1)}$$

$$negativity = anger + sorrow - joy - surprise \quad \text{(equation 2)}$$

where each emotion in the equation means the numeric level of possibility of this emotion. I use two different measures, as I find that it is difficult to place "surprise" on a scale from negative to positive. Many previous studies (e.g., Mohammad & Turney, 2013) find "surprise" is more associated with positive sentiment; however, I manually checked the ad images that are annotated as "very likely" to evoke surprise and find that most of the ad images with the highest "surprise" score seem to be negative ads. Some examples are shown in Appendix (7); the two images are displayed very frequently with a high level of CPM, but they seem to evoke fear and disgust rather than surprise; however, fear and disgust are not included in the emotion classifier. Therefore, including "surprise" as a factor of positive sentiment may be misleading. As a result, I include both measurement methods in the first model to measure the sentiment score for a robustness check, and the models are shown below:

$$eCPM = \beta_0 + \beta_1 negativity + \beta_2 t + \beta_3 party + \beta_4 thirdparty + \mu$$

$$eCPM = \beta_0 + \beta_1 anger + \beta_2 joy + \beta_3 sorrow + \beta_4 surprise + \beta_5 t + \beta_6 party + \beta_7 thirdparty + \mu$$

4 More details about Google Cloud Vision: https://cloud.google.com/vision/overview/docs/

However, due to the different sets of emotions used in classifying text ads and image ads, it is difficult to compare or conclude the effects of some emotions in these two types of ads. Google Cloud Vision classifies fewer emotions than the NRC lexicon; only anger, joy and surprise are captured by both classifiers. Thus, Google Cloud Vision may involve other uncovered emotions (but covered by the NRC lexicon for text ads) in the 4 emotions it can detect, such as the case of surprise mentioned above. The emotion of "sorrow" in the image ads is not equal to "sadness" in the text ads data set, as it also seems to include other negative emotions, such as disgust, that tend to have different effects on the CPM. Nonetheless, two basic emotions of anger and joy annotated in both the text and image ads are still comparable.

To test the overall effect of the interested emotion on all ads, I first normalize the emotion scores in the two types of ad datasets to 0-1. A dummy that indicates whether an ad is a text or image is included, and then an interaction term between this type dummy and the emotions score is also included. In this way, any difference in the scale of the variable that affects the CPM would be captured by the interaction effect. This interaction effect also tests whether emotions have different effects depending on whether they are conveyed by the text or the image.

$$eCPM = \beta_0 + \beta_1 emotion + \beta_2 type + \beta_3 emotion * type + \beta_4 t + \beta_5 party + \beta_6 thirdparty + \mu$$

Furthermore, I investigate if the candidate used provocative content with strong anger to attract voters and stimulate the engagement-based impression algorithm.

**4.2 Topic models and keyness study**

I apply topic models to classify the topics in each text for investigating questions about what topics political advertisers mention in electoral ads and whether particular topics are more likely to be negative. The most commonly used topic model is Latent Dirichlet Allocation (LDA). However, the LDA model may not work well for short texts. The technical reason is that these traditional topic models, such as LDA, capture the latent patterns of word co-occurrence at the document level to classify topics, and they may not work well due to the shortage of words in short documents. The ad content data I collected is a typical short text, as each of the ads only contain a title and a main body, with one or two sentences (see appendix 3). To classify the topic validity, I introduce the BTM to classify the topics. BTM is a newly developed topic for short text, especially, the text in social media (Yan el., 2013), which is based on word co-occurrence that learns topics by modelling the word-word co-occurrence patterns (for example, biterms) in the whole corpus. As a result, I use the relational biterm topic model to classify the topics in this dataset and then use LDA for a robustness check.

For the image ads, I study the textual labels from the image content detected by AI (as shown in appendix 5)and apply keyness analysis (Dunning, 1993; Bondi & Scott, 2010) to investigate what content in the images is correlated more with the higher negativity that the negative advertisers prefer to use.

**5. Results:**

**5.1 Data Visualization of Political Ads on Google:**

After calculating real score values of the emotions and sentiments in both text ads and image ads, I classify the sentiment in each ad according to its sentiment score. For the text ad, a positive score and a negative score are compared to select the higher one as the label; if two sentiment values are the same, the ad is labelled as “Neutral”. For the image ad, the classification is based on a negativity value (equation 1), as 0 means “Neutral”, higher than 0 is labelled as “Negative”, and lower than 0 is labelled as “Positive”, accordingly. The preliminary results can be seen in the figures below. The clear difference in sentiments can be seen in Figure 4, with image ads having a much larger share of neutral ads and a smaller share of positive ads. It is also worth noting that the share of the image ads going negative is only approximately half of that of the text ads (8.51% vs. 17.55%). Overall, negative ads have the smallest percentage in both types of ads, indicating that negative political advertising is not dominant but cannot be ignored.

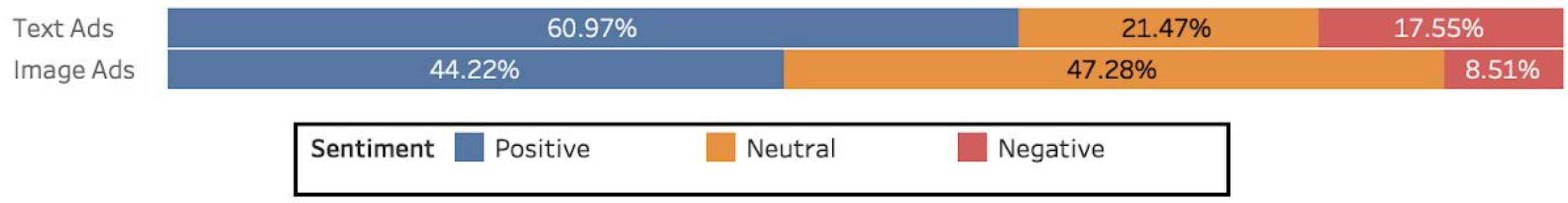

Figure 4: Distribution of Political Ads By Sentiments on Google

To study the pattern of emotions, I classify the major emotion in each ad. The method searches for the emotion with the highest real-valued score (text ads) or probability (image ads) for each ad and then labels it as the major emotion. For cases where more than one emotions with the highest value, I label it as “Mixed”. Figure 5 and Figure 6 show the distribution of major emotions together with the sentiments for textual ads and image ads, respectively. As I discussed previously, the large difference in shares also suggests that the classifications of emotions are not very consistent. For instance, “Trust” occupies the highest percentage, i.e., 60.97%, in text ads, which is not even covered by the classifier of image ads. In comparison, the share of “joy” in the image ads (47.28%) s much higher than that of the text ads (0.73%); the extremely small share of “joy” in text ads (0.73%) also indicates that joy may be replaced by other positive emotions, such as trust, in the text ads compared to image ads. However, the shares of my interested emotion, i.e., anger, in both types of ads are very close (2.88 vs. 2.2%), implying that classifying anger is very consistent and uncontroversial.

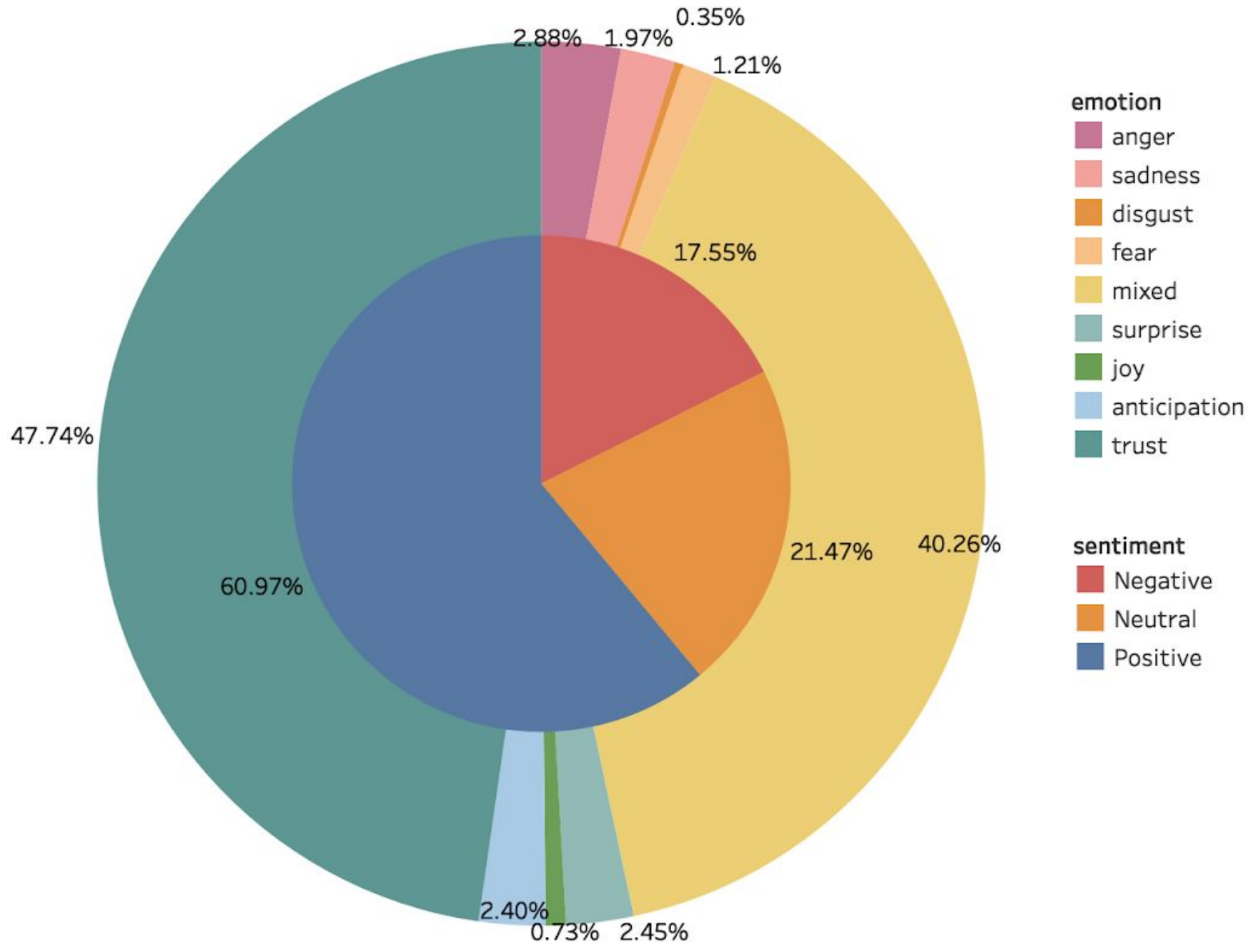

Figure 5: Distribution of Textual Political Ads By Sentiments and Emotions

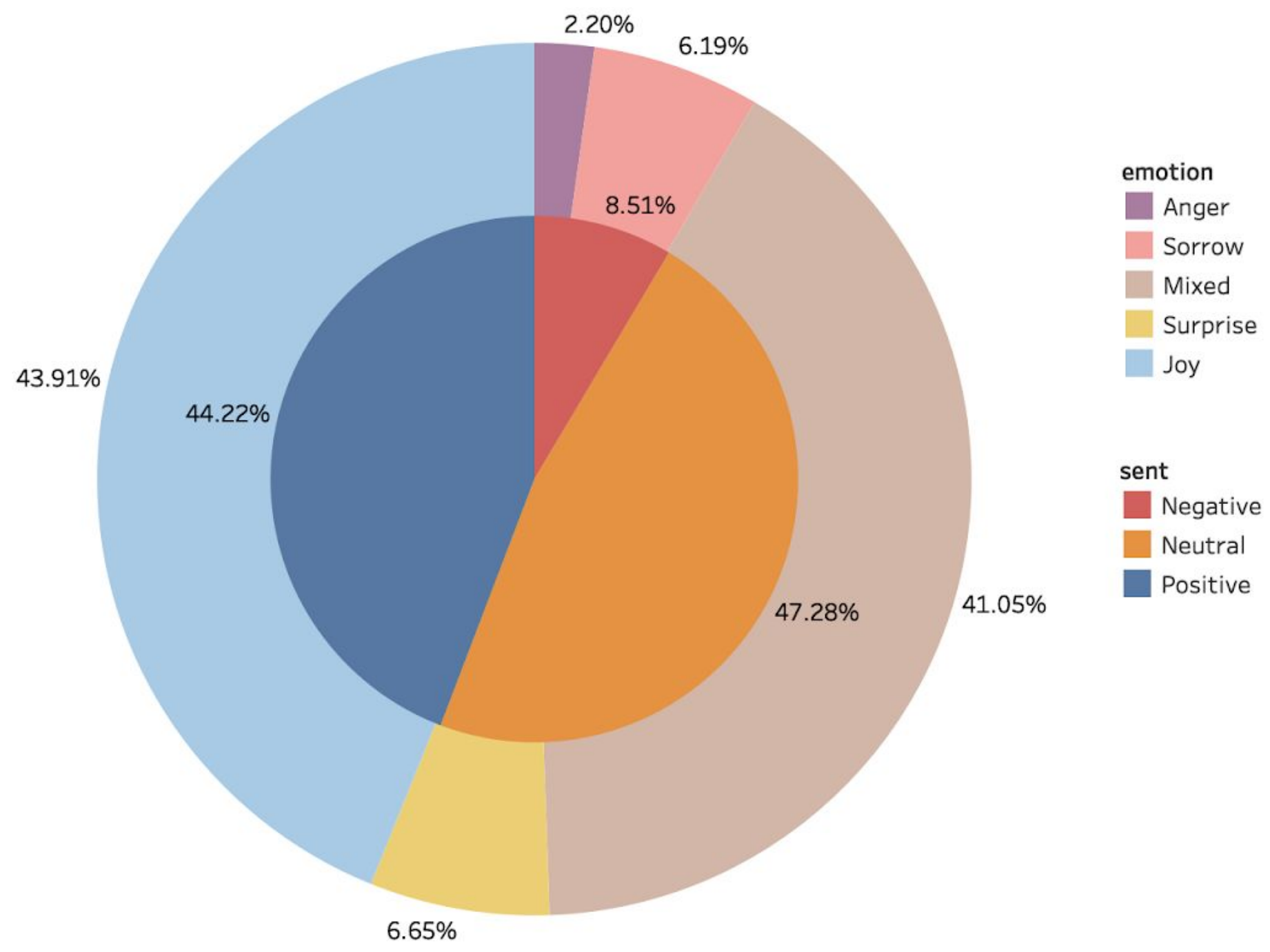

Figure 6: Distribution of Image Political Ads By Sentiments and Emotions

Furthermore, I also investigate the geographic distribution of negative ads. However, most of the negative ads detected are from the third party organizations (as shown in Figure 5), which cannot be matched with the regions in most of cases. As a result, only the negative ads from the candidates and some local third party organizations are included in Figure 7. Interestingly, the negative ads are not focused on the larger states with more elections, as the top 3 with the most negative ads are Virginia, Massachusetts and Florida instead of California (rank 7) or Texas (rank 6), where the local populations were huge and elections were more and competitive. Although this map only included the negative ads mainly from candidates and lacks precise data of impressions, it still suggests that the distribution of negative ads has a pattern that not only depends on the number of elections and population but may likely depend on other factor such as the intensity of the local races.

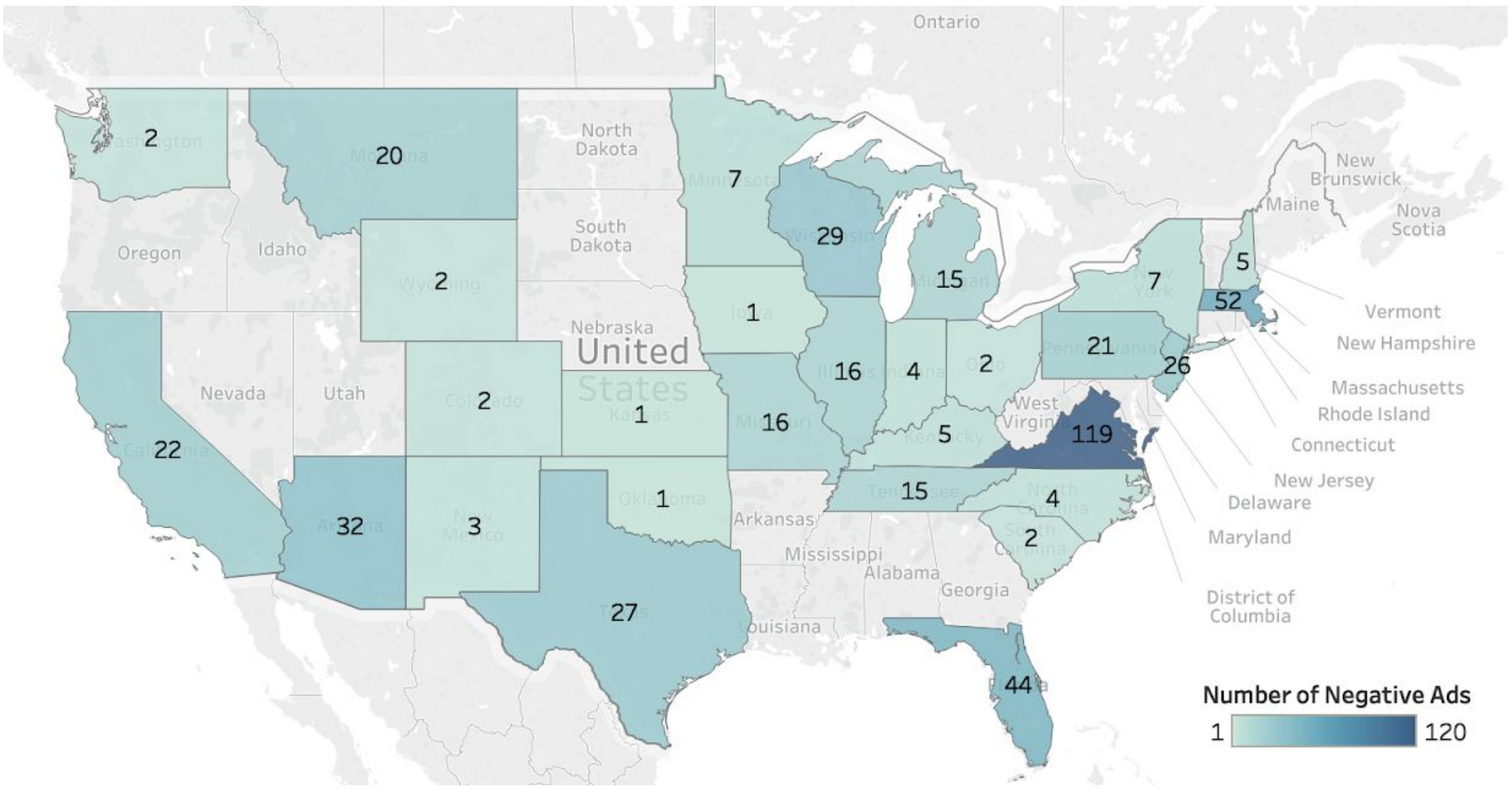

Figure 7: Distribution of Negative Political Ads By State

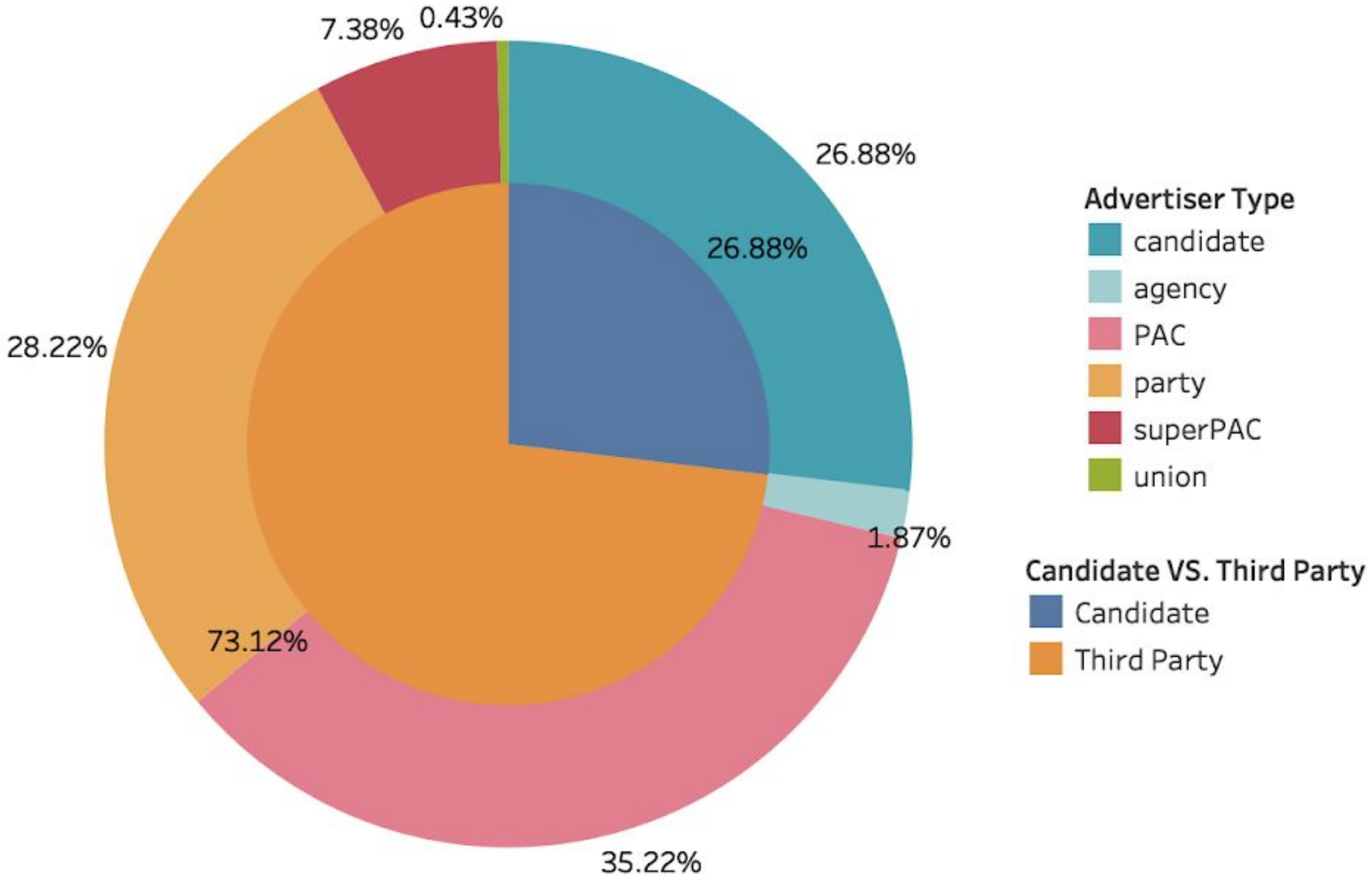

Figure 8: Distribution of Negative Ads by Advertiser Types

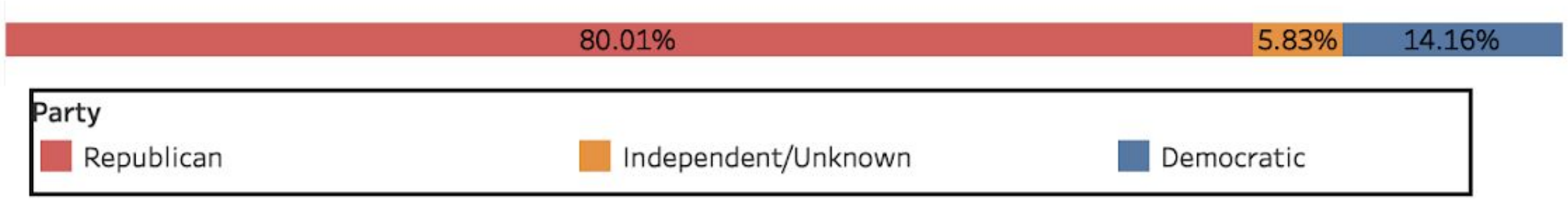

Figure 9: Distribution of Negative Ads by Party

To answer the second descriptive question about who are more likely to use negative advertising strategy, I visualize the share of total negative ads in figure 8, figure 9 and display the top 10 advertisers' basic information by table 1. As seen clearly from figure 8, the third party organizations delivered approximately three quarters of the total negative ads during the midterm elections, which is much more than that of candidates. Within the third party organizations, PAC contributes the most (35.22%), while the political party's share is the second highest (28.22%). In contrast, candidates only delivered 26.88% of the negative ads.

| Advertiser Name | Number of Negative Ads | Number of Ads | Negative Ads' Share | Advertiser Type | Party |
|---|---|---|---|---|---|
| REPUBLICAN NATIONAL COMMITTEE | 517 | 523 | 98.85% | party | R |
| NATIONAL HORIZON | 291 | 1438 | 20.24% | PAC | R |
| SENATE LEADERSHIP FUND | 191 | 1512 | 12.63% | PAC | R |
| COMSTOCK FOR CONGRESS | 103 | 174 | 59.20% | candidate | R |
| PRIORITIES USA ACTION & HOUSE MAJORITY PAC | 80 | 304 | 26.32% | superPAC | D |
| RESTORATION PAC | 48 | 184 | 26.09% | PAC | R |
| DIEHL FOR US SENATE | 40 | 390 | 10.26% | | R |
| DEDICATEDEMAILS.COM | 35 | 90 | 38.89% | agency | |
| AMERICANS FOR PROSPERITY | 31 | 181 | 17.13% | PAC | R |
| MCSALLY FOR SENATE INC | 27 | 352 | 7.67% | candidate | R |

Table 1: Table of Top 10 Advertisers by Size of Negative Ada

Table 1 shows the top users of negative advertising. The most striking result is the first advertiser type, i.e., the "Republican National Committee"; this advertiser runs the most negative ads (517) and uses ads that are almost all negative (negative ad share: 98.85%). In fact, the Republican National Committee leads the Republican Party nationally, especially regarding election arrangements, and can reflect the Republican attitude in the campaign.

Figure 9 also suggests this, as 80.01% of the negative ads are from Republican advertisers. Additionally, 8 out of the top 10 negative advertisers are Republican, implying that Republicans are more likely to use negative advertising and more frequently. Only 1 negative advertiser is a Democrat, i.e., "PRIORITIES USA ACTION & HOUSE MAJORITY PAC", and it only has 80 negative ads and 26.32% of the negative ad share. Furthermore, there are only 2 advertisers that are Republican candidates; the others are all third party organizations, confirming that most negative advertising is performed by third party organizations. The candidate that used the most negative ads in the midterm election is "COMSTOCK FOR CONGRESS", the official campaign team of Barbara Comstock, who was the Republican candidate for Virginia's 10th congressional district election.

### 5.2 Results of the Topic Models and Keyness Analysis

For the last descriptive question, I apply the topic model (BTM) to determine what topics the political advertisers typically use in text ads for negative advertising. I select K = 30, as the topic model is the most fit and valid at K = 30 (see appendix 8 for all topics with top terms). A clear variation of sentiments across different topics can be seen from figure 10; the highest percentage of negative ads is approximately 80%, while the lowest is near zero. I label all the topics manually and find that topic 12 has the highest share of negative ads, which involves attacks on Jennifer Wexton, the Democratic candidate of Virginia's congress election, regarding her policy on tax-raising. The second highest one is topic 13, which is about attacking Mark Harris, the Republican candidate for North Carolina's congress election, for his attempt to cut healthcare. The third topic attacks the Democratic senate candidate Tammy Baldwin for her tax-raising policy. This is not too surprising that the Republicans often attack Democrats on taxes and Democrats often attack Republican for cutting health care. This result also confirms the findings of frequent negative ads in Virginia, Wisconsin and North Carolina in figure 7. However, this result does not live up to my expectation, as the topic models focused too much on the specific candidates rather than the general policy topics.

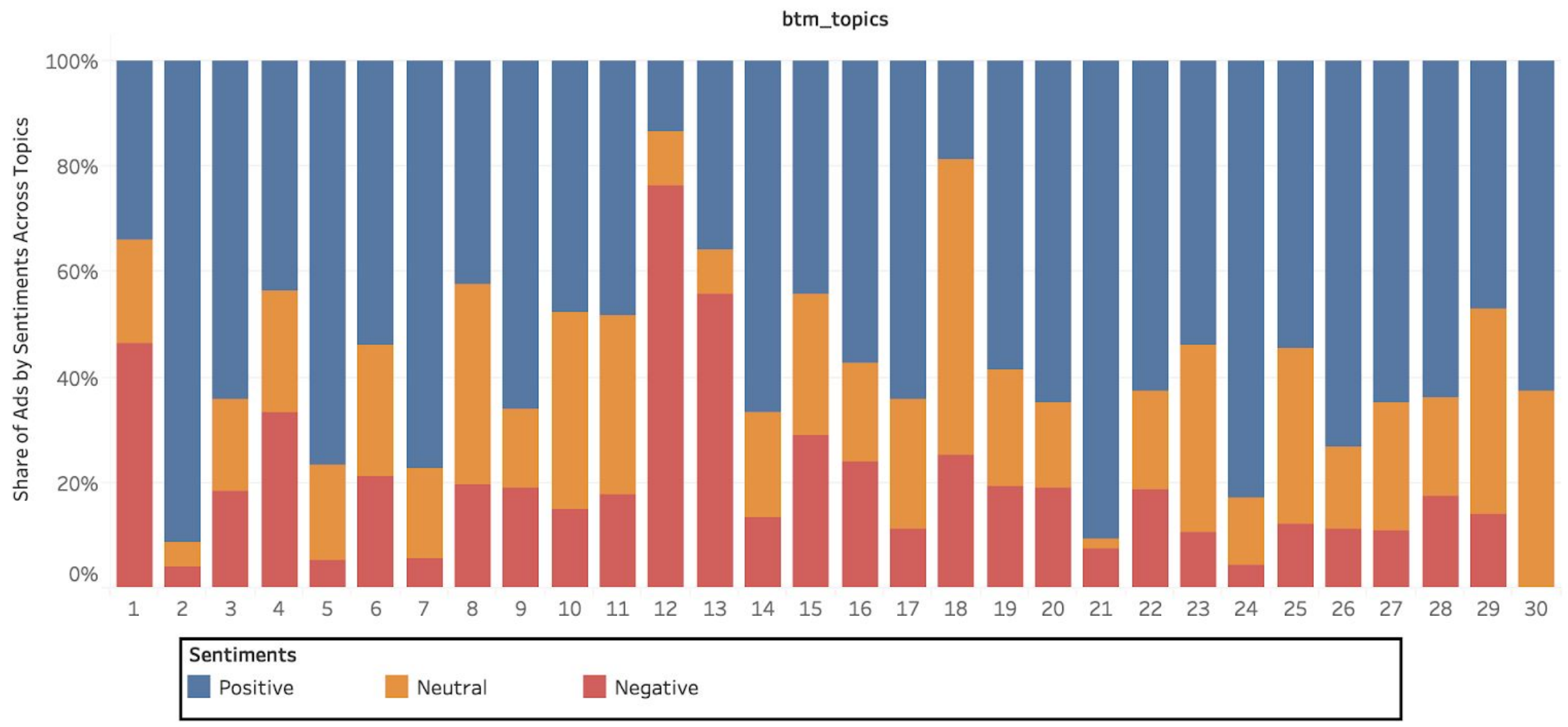

Figure 10: Distribution of Sentiments in Topics by BTM

For a robustness check, I also apply the LDA topic model, as shown in figure 8. The LDA also gives a very similar result. Topic 1, with approximately 80% negative ads, is again about attacking Jennifer Wexton for her policy on tax-raising, while topic 2 is still about attacking Mark Harris for healthcare. As a result, the topic model results should be robust.

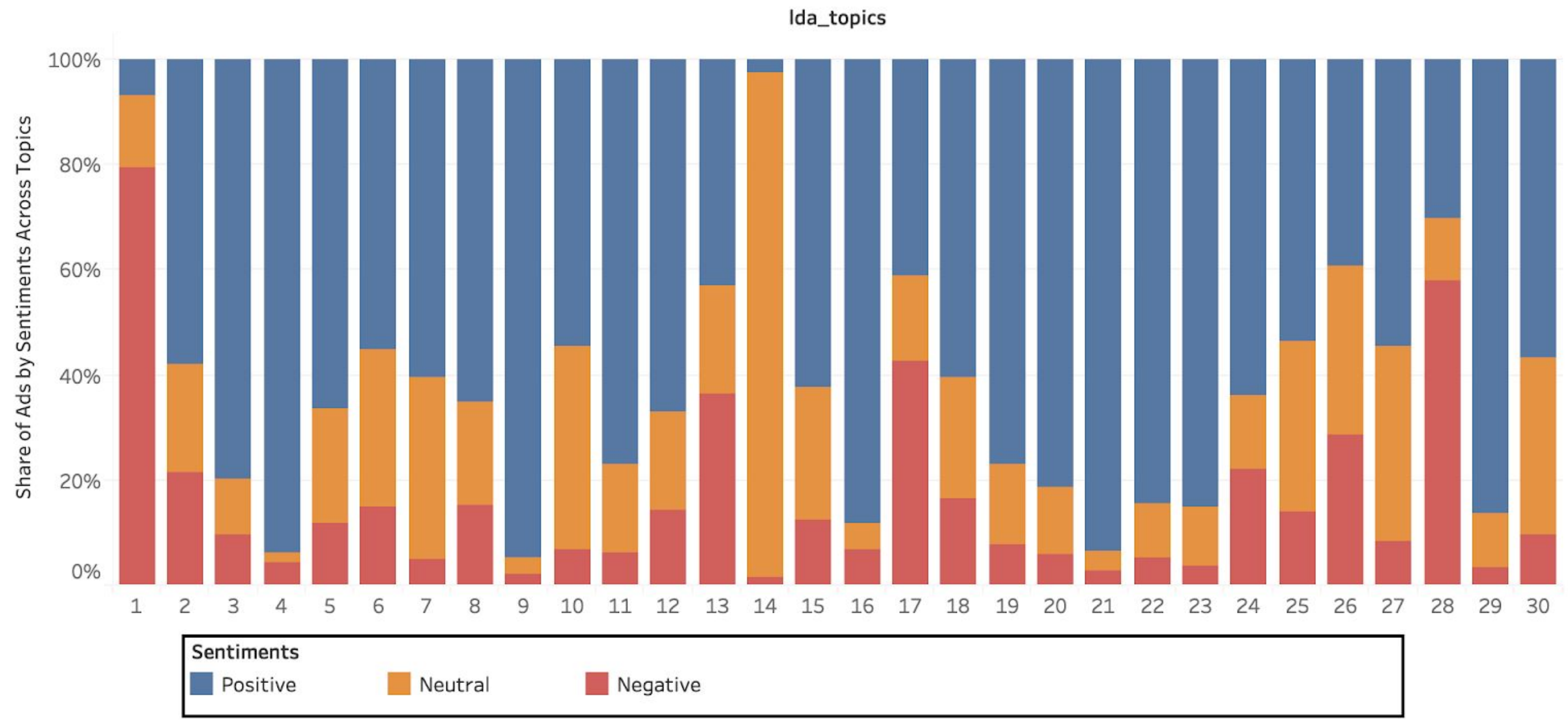

Figure 11: Distribution of Sentiments in Topics by LDA

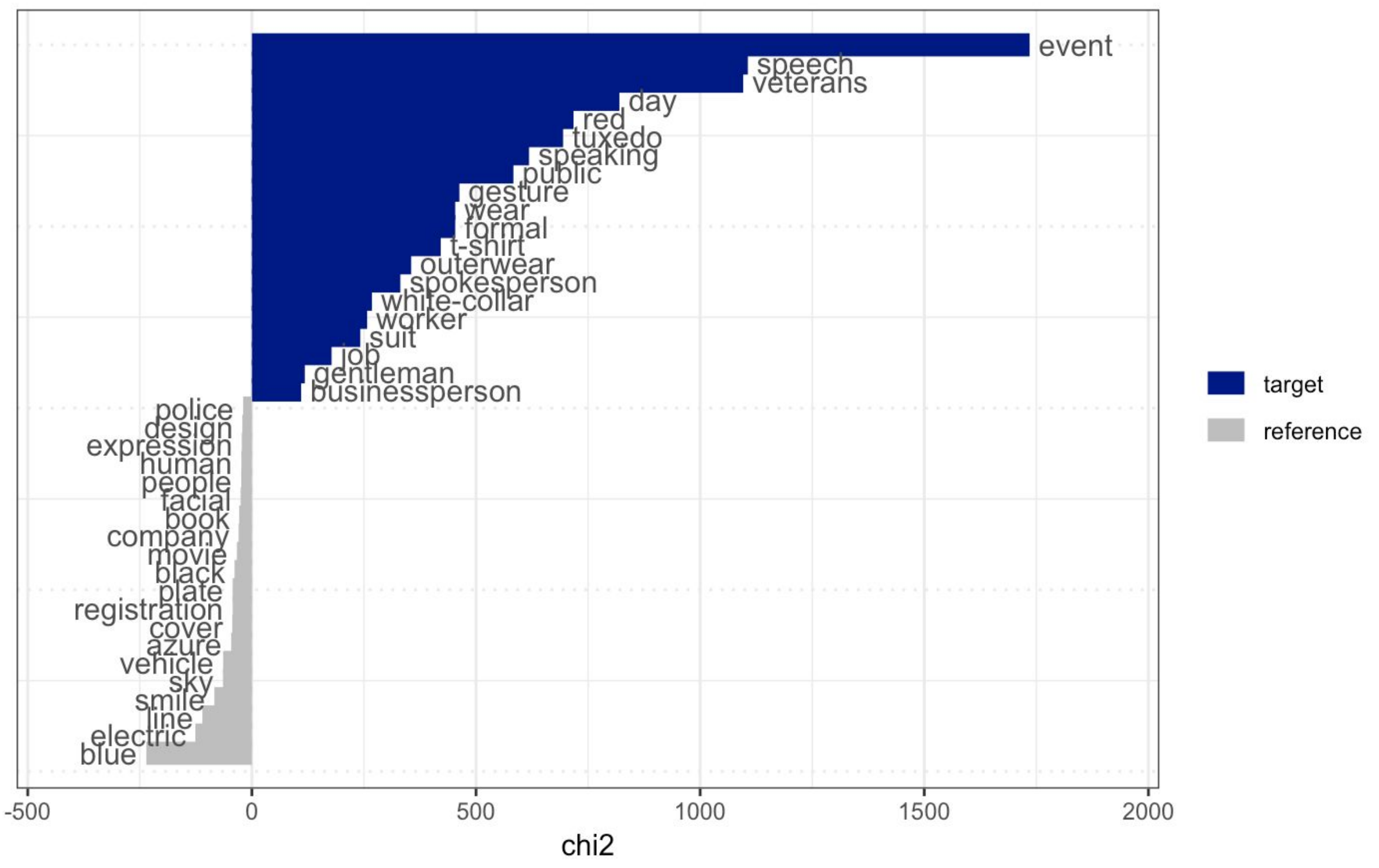

Figure 12: Keyness Analysis of the Negative Ads and Non-negative Ads

For the image ads, although the words can be detected, its textual content is difficult to apply to topic models, as the text in the image ads is very short and fragmented. Additionally, most of the text in the image ads is the information disclosure and slogans instead of key topics. However, the key content of images can be detected, as there are many labels (detected by AI from Google Cloud Vision) that can be used to study what type of content the advertisers use to evoke negativity. Keyness analysis is then applied to investigate what labels are more likely to be negative, after removing stop words and repeated words of facial features (as there are many faces in the ads that can be detected as many facial organs). The results are shown in figure 9, which suggests that the keyness labels of the labels are quite different in these two groups. The first key difference is about colour; the negative ads are more likely to use red, while the non-negative ads are more likely to use blue (includes azure). This may be because red implies the strong negative emotions of anger or fear and wrongness, while blue implies joy, peace and rightness. Regarding the specific aspects that make the two sides different, one clear finding is that negative ads include more formal settings, such as “events”, “speeches”, “tuxedos”, “public”, “formal”, “spokesman”, and “gentleman”, which seems to be distant from normal voters. In contrast, the non-negative ads include more content that are close to the normal people, such as “electric”, “smile”, and “plate”. Another key difference is that the negative ads include the words “job”, “white-collar” and “worker”, which are related to the hot debate of employment in the elections. This may suggest that “employment” is a good point to attack the rival’s policy.

**5.3 Ordinal Logistic Regression Results:**

The most important question that about the mechanism of the negative advertising affect voters’ behaviours is tested with ordinal logistic regression models as follows.

Ordinal Logistic Regression Result (Text Ad)

| | *Dependent variable:* | |
|---|---|---|
| | eCPM | |
| | (1) | (2) |
| Negativity | −0.005 | |
| | (0.043) | |
| Anger | | −0.318*** |
| | | (0.082) |
| Sadness | | 0.058 |
| | | (0.088) |
| Joy | | 0.238*** |
| | | (0.088) |
| Fear | | 0.239*** |
| | | (0.072) |
| Anticipation | | 0.214*** |
| | | (0.074) |
| Disgust | | −0.115 |
| | | (0.079) |
| Surprise | | −0.003 |
| | | (0.081) |
| Trust | | −0.024 |
| | | (0.042) |
| Time Span | 0.010*** | 0.010*** |
| | (0.002) | (0.002) |
| Third−party | 0.137 | 0.084 |
| | (0.094) | (0.100) |
| Party(Independent/Unknown) | −1.035*** | −1.057*** |
| | (0.143) | (0.148) |
| Party(Republican) | 0.336*** | 0.277*** |
| | (0.083) | (0.089) |
| Observations | 3,959 | 3,959 |
| *Note:* | *p<0.1; **p<0.05; ***p<0.01 | |

(Table 2)

Ordinal Logistic Regression Result (Image Ads)

| | *Dependent variable:* | | |
|---|---|---|---|
| | eCPM | | |
| | (1) | (2) | (3) |
| Negativity | −0.052*** | | |
| | (0.011) | | |
| Negativity(Equation2) | | −0.050*** | |
| | | (0.011) | |
| Anger | | | −0.200** |
| | | | (0.097) |
| Sorrow | | | −0.752*** |
| | | | (0.063) |
| Joy | | | 0.009 |
| | | | (0.012) |
| Surprise | | | −0.073** |
| | | | (0.037) |
| Party(Independent/Unknown) | −0.506*** | −0.496*** | −0.527*** |
| | (0.078) | (0.078) | (0.079) |
| Party(Republican) | −0.672*** | −0.666*** | −0.628*** |
| | (0.067) | (0.068) | (0.068) |
| Third−party | 0.173*** | 0.165*** | 0.218*** |
| | (0.053) | (0.053) | (0.054) |
| Time Span | −0.021*** | −0.021*** | −0.021*** |
| | (0.001) | (0.001) | (0.001) |
| Observations | 13,822 | 13,822 | 13,822 |
| *Note:* | *p<0.1; **p<0.05; ***p<0.01 | | |

(Table 3)

In Table 2, column 1 shows that negativity score of ad content is negatively related to eCPM but is not significant, controlling the time span of the ad's display, partisan and advertiser type. This may indicate that aggregating emotions to a sentiment scale of negativity is not appropriate, as different emotions may have different effects on the eCPM. Furthermore, I investigate each specific emotion rather than overall negativity and find that the two negative emotions, i.e., anger and fear, both have significant effects on the eCPM; however, the signs are different. The anger in text ads seems to reduce the eCPM, while an increase in fear increases the eCPM. In other words, the more anger the ads show, the more likely that the audience will click the ads to view the additional information that the advertisers want to diffuse, while the more fear in the ads will discourage people from clicking on the ad. This result aligns with Ryan's finding (2012), which suggest that the anger in text ads can stimulate people's political information seeking and then change people's political behaviours.

In Table 3, I establish three versions of the ordinal regression model to study the image ads by including different sets of measures of negativity and emotions, and control the other variables. Column (1) excludes "surprise" and only takes joy as a positive emotion (equation 1), while the column (2) calculates the negativity of image ads by including surprise as a positive sentiment (equation 2). Both measures shows a similar result, i.e., that negativity has a clear negative effect on eCPM, indicating that campaigns with more negative content can significantly improve engagement and, thus, reduce eCPM (Hypothesis 1). This finding suggests that the more negative political ads can better draw audiences' attention and make audiences more likely to engage. Column 3 shows the result from the perspective of specific emotions; the negative emotions of anger and sorrow have significantly negative effects on the eCPM, and joy has a positive effect on the eCPM, but it is not significant. The result of anger clearly aligns with the text ads, as expected.

**Ordinal Logistic Regression Result (Both Types of Ads)**

| | *Dependent variable:* |
|---|---|
| | eCPM |
| Anger | $-0.525^{**}$ |
| | (0.209) |
| Type(Image) | $-2.847^{***}$ |
| | (0.071) |
| Anger*Image | 0.176 |
| | (0.357) |
| Time Span | $-0.012^{***}$ |
| | (0.001) |
| Third-party | 0.004 |
| | (0.045) |
| Party(Independent/Unknown) | $-0.397^{***}$ |
| | (0.062) |
| Party(Republican) | $-0.337^{***}$ |
| | (0.051) |
| Observations | 17,781 |
| *Note:* | $^{*}$p<0.1; $^{**}$p<0.05; $^{***}$p<0.01 |

(Table 4)

Finally, both the text and image ads datasets are included in the model to compare and test again to see if the effect of anger on engagement is consistent and robust. The result aligns with the previous tests, confirming that anger has a significant negative impact on the eCPM, i.e., political ads containing anger should be associated with a lower eCPM, implying that the more anger a political ad evokes, the higher audience engagement level it evokes. Furthermore, the interaction term in the model is not significant, suggesting that the effect of anger on the eCPM is consistent for both types of ads.

### 5.4 Further Robustness Checks:

The above tests may trigger some doubts; the first is whether the possibility level of the emotions in the image dataset should be treated as a continuous measure. As this measure is categorical in nature, the distance between "very unlikely" and "likely" may not be equal to the distance between "possible" and "likely". Regardless, to ensure that the numeric measure is robust, I also test it as a categorical variable.

Multinomial Logistic Regression Result(Image Ad)

| | *Dependent variable:* |
|---|---|
| | eCPM |
| AngerUNLIKELY | −0.179 |
| | (0.161) |
| AngerPOSSIBLE | −0.489* |
| | (0.270) |
| AngerLIKELY | −0.418 |
| | (0.727) |
| SorrowUNLIKELY | −0.882*** |
| | (0.079) |
| SorrowPOSSIBLE | −1.021** |
| | (0.435) |
| SorrowLIKELY | −1.509*** |
| | (0.346) |
| JoyUNLIKELY | −0.119 |
| | (0.102) |
| JoyPOSSIBLE | 0.114 |
| | (0.162) |
| JoyLIKELY | 0.627*** |
| | (0.132) |
| JoyVERY_LIKELY | −0.004 |
| | (0.049) |
| SurpriseUNLIKELY | 0.101 |
| | (0.128) |
| SurprisePOSSIBLE | −0.033 |
| | (0.163) |
| SurpriseLIKELY | −0.114 |
| | (0.208) |
| SurpriseVERY_LIKELY | −0.547*** |
| | (0.197) |
| thirdparty | 0.203*** |
| | (0.055) |
| t | −0.021*** |
| | (0.001) |
| PartyIndependent/Unknown | −0.513*** |
| | (0.081) |
| PartyR | −0.606*** |
| | (0.069) |
| Observations | 13,822 |
| *Note:* | *p<0.1; **p<0.05; ***p<0.01 |

(Table 5)

This model treats the possibility levels of the emotions in the image ads as discrete categorical values instead of continuous numeric values, and also indicates a very similar result to the previous model. It is clear that the categories are all increasing or decreasing in similar magnitude, so in this case, it should be justified to treat it as a numeric variable. Additionally, as I find above, the signs of the coefficients of “unlikely” and other higher levels of surprise are opposite; however, only the effect of the "very likely" level of surprise is significantly negative. As shown above, this indicates that some of the most repeated ads at the "very likely" level of surprise involve other emotions other than the "surprise" detected by AI. This result can justify the exclusion of surprise in the calculation of sentiment and in the comparison.

OLS Regression Result (Text Ad)

| | Dependent variable: | |
|---|---|---|
| | eCPM | |
| | (1) | (2) |
| Negativity | −0.001 | |
| | (0.031) | |
| Anger | | $-0.215^{***}$ |
| | | (0.056) |
| Sadness | | 0.039 |
| | | (0.064) |
| Joy | | $0.157^{**}$ |
| | | (0.062) |
| Fear | | $0.182^{***}$ |
| | | (0.053) |
| Anticipation | | $0.155^{***}$ |
| | | (0.054) |
| Disgust | | −0.088 |
| | | (0.055) |
| Surprise | | −0.004 |
| | | (0.058) |
| Trust | | −0.019 |
| | | (0.029) |
| Time Span | $0.007^{***}$ | $0.007^{***}$ |
| | (0.001) | (0.001) |
| Third−party | 0.094 | 0.042 |
| | (0.068) | (0.071) |
| Party(Independent/Unknown) | $-0.566^{***}$ | $-0.570^{***}$ |
| | (0.087) | (0.090) |
| Party(Republican) | $0.261^{***}$ | $0.210^{***}$ |
| | (0.061) | (0.065) |
| Constant | $5.761^{***}$ | $5.787^{***}$ |
| | (0.062) | (0.074) |
| Observations | 3,959 | 3,959 |
| $R^2$ | 0.039 | 0.049 |
| Adjusted $R^2$ | 0.037 | 0.046 |
| Residual Std. Error | 1.648 (df = 3953) | 1.641 (df = 3946) |
| F Statistic | $31.832^{***}$ (df = 5; 3953) | $16.950^{***}$ (df = 12; 3946) |
| *Note:* | | $^{*}$p<0.1; $^{**}$p<0.05; $^{***}$p<0.01 |

(Table 6)

OLS Regression Result

| | Dependent variable: | |
|---|---|---|
| | eCPM | |
| | (1) | (2) |
| Negativity | $0.033^{***}$ | |
| | (0.008) | |
| Anger | | $-0.138^{**}$ |
| | | (0.070) |
| Sorrow | | $-0.584^{***}$ |
| | | (0.054) |
| Joy | | 0.005 |
| | | (0.008) |
| Surprise | | $-0.045^{*}$ |
| | | (0.026) |
| Time Span | $-0.016^{***}$ | $-0.016^{***}$ |
| | (0.001) | (0.001) |
| Third−party | $0.204^{***}$ | $0.227^{***}$ |
| | (0.038) | (0.039) |
| Party(Independent/Unknown) | $-0.355^{***}$ | $-0.372^{***}$ |
| | (0.054) | (0.054) |
| Party(Republican) | $-0.489^{***}$ | $-0.459^{***}$ |
| | (0.045) | (0.045) |
| Constant | $6.688^{***}$ | $7.485^{***}$ |
| | (0.044) | (0.116) |
| Observations | 13,822 | 13,822 |
| $R^2$ | 0.032 | 0.039 |
| Adjusted $R^2$ | 0.031 | 0.039 |
| Residual Std. Error | 1.710 (df = 13816) | 1.703 (df = 13813) |
| F Statistic | $90.285^{***}$ (df = 5; 13816) | $70.446^{***}$ (df = 8; 13813) |
| *Note:* | | $^{*}$p<0.1; $^{**}$p<0.05; $^{***}$p<0.01 |

(Table 7)

Another issue is whether it is appropriate to treat eCPM as an ordered categorical variable, as eCPM in the real world is a continuous measure. To address this concern, OLS models are also established to replicate the tests above, which treat the estimated eCPM as a continuous number. The results consistently confirm the previous findings, indicating that the negativity of ads on eCPM is not clear for text ads, but it is significant and constant for image ads; anger is also negatively correlated with eCPM and very similarly to the other variables.

**6. Discussion**

First, the strongest finding across these empirical models confirms the consistent impact of anger on improving engagement in terms of clicks and information-seeking, which is a sound explanation of how negative ads work. Consider the substantive significance of the effect of anger on people's engagement. If the score of anger increases 1 unit in ads, the eCPM is 27% and 18% more likely to decrease to a lower level for text ads and image ads, respectively, controlling for other variables. This amount of the effect is vital enough for the candidates in an election, which suggests that political advertisers have a strong incentive to use highly emotional ads and negative ads in campaigns to attract and then persuade voters with the help of the engagement-based ad distribution model by Google Ads. In particular, it is evident that the anger in negative ads strongly evokes engagement, which is a logic mechanism of how negative advertising works.

Second, the empirical results show that the effect of negativity is mixed. For the text ads, there is no evidence that there is an effect of negativity on engagement. For the image ads, although I capture a significant effect of negativity on improving engagement, it is far weaker than the effect of anger alone. Additionally, it is worth noting that I use the possibility level of emotions to calculate the negativity for image ads, so that this significant effect of negativity here is also driven by anger. Thus, I cannot find substantial evidence for the negative bias theory in explaining the mechanism of negative advertising.

Third, the empirical results also show that Republican ads are better than those of Democrats in the measure of eCPM, controlling for other variables. The time span is very consistent and robust, as expected, which is most likely to indicate that political advertisers select and then keep the outperformed ads with a lower eCPM running for a longer time, and discontinue use of the other ads with high eCPM. The effect of third-party organizations on eCPM seems to be positive but only significant for image ads. This is likely because the third-party organizations are required by the new law to declare that they are sponsored by a third-party by stating that "the ad is not authorized by any candidate" (as shown in the text cloud of detected text from image ads) in the image ads, which may alarm the audiences of the intended negative advertising. And they usually do not include such statements in the text ads, so that audience may not be that alerted.

Fourth, to investigate the content in the ads, I apply a newly developed topic model (BTM) to investigate the topics of the text ads and their relationship with negative advertising. I find that the topics that the two parties use in their negative advertising are the topics they discuss in public, such as taxes and healthcare. Regarding the image ads, I apply keyness analysis to the textual labels that are detected from the images. I find that the images created are also very different between negative and non-negative ads, as both the negative ads and non-negative ads prefer some particular key features in terms of colours, topics, and distance with voters.

Finally, I first include sentiment analysis on the image data to study political ads on social media. The descriptive results show that the share of negative ads is still the minority but should not be ignored. I also find that Republican advertisers and the third party organizations used negative ads more frequently in these midterm elections. The geographic distribution of negative ads very likely depends on the intensity of races, as the negative ads tend to appear more in those states with close races.

**7. Implication:**

I propose a novel approach to classifying negative image ads and calculate the negativity based on the emotions evoked from the ad content through image recognition technology. As far as I know, this is the first time that AI image recognition technology is applied to studies in political science.

The results above contribute to the study of political communication and political psychology in several ways. First, this study contributes to the limited literature about how negative ads influence voter behaviour, and it contributes by revealing the patterns of political campaigns on social media through a big data approach. Second, it also contributes to the growing empirical studies regarding the effect of political ads in digital platforms on voter behaviour (Spenkuch & Toniatti, 2018; Bond et al., 2012; Liberini et al., 2018).

This study is also valuable for policy-making and practice. Many empirical findings have shown that the political campaigns on social media have significant but unwanted effects on voter behaviour, and even on the democratic system, as a result of the sharp rise of online populism recently in the democratic countries, especially at the time when regulations were absent. The findings in this research will provide unprecedented insight about the negative campaign strategies used, which can help to better regulate the campaign policies to prevent unwanted consequences in elections and unwanted failures of democracy.

For the general public, this study may help people become aware of the effects of political ads on social media as well as the “tricks” of the political advertisers, which may guide them to vote more rationally in the future elections.

**8. Limitations and Further Research:**

The first limitation is that our text ads’ content dataset is not complete and may be biased, as Google has blocked 39% of the textual ads that violated the Google ad policy, which may include a large number of negative ads.

The second limitation is that image recognition may not work well for images without a face (most images do include faces) according to Google. As the image recognition model provided by Google first searches for faces in the image to analyse the emotions and then other non-human features that may suggest emotions. The future research can develop a trained machine learning model to classify the negative ads by more features.

The last limitation is that using sentiment to classify and scale negative advertising may not be precise. For instance, my sentiment prediction based on the dictionary is not able to identify all the negative textual ads, as some of the negative advertising only uses non-negative words in content but directs the audience to an external website containing negative messages. This intentional manipulation involving using non-negative words to perform negative advertising may be a way that the political advertisers use to avoid being blocked by Google, causing our words-based sentiment classifier to capture fewer negative ads, which can lead to a biased result in the empirical study. This may be the reason that the effect of negativity is only captured by the image ads dataset. To address this problem, future research can develop a machine learning model to classify the negative ads rather than using the sentiment to capture negative ads, and may also make a new scale to measure negativity in textual ads.

## 9. Conclusion

This research studies the overall patterns of increasingly widespread negative campaigns on the social media in the elections and finds out the working mechanism of how negative political ads on social media affect voter behaviour with a set of novel methodologies. With the sentiment analysis, this research find that negative ads are still the minority in general but are popular in some particular areas with intensive races. The Republican and third party organizations are found to use more negative ads. Furthermore, I also find that the some particular content in ads are more related to negative ads. The topics in the text political ads that are more likely to go negative are still these debatable topics such as health care and tax, while the content in negative image ads use more red color and formal settings.

A growing literature (i.e. Gordon & Hartmann, 2013; Chung & Zhang, 2016; Ceron & Adda, 2016; Wang, Lewis & Schweidel, 2018) has shown that political advertising as well as negative advertising is effective in changing voters' behaviour, this study shed light on the neglected mechanism of how the increasing negative ads affect voters' behaviour. By a set of empirical tests, I observe a strong correlation between the degree of anger evoked and the engagement measure (eCPM). It can be concluded that the anger evoked information-seeking is one of the main channel that make negative ads effective rather than the commonly believed mechanism introduced by negative bias theory.

**Acknowledgments**

The author wishes to thank Pablo Barberá, Blake Miller for their valuable helps, and Department of Methodology of London School of Economics and Political Science for providing financial support for this study.

**Referemce:**

**Appendix:**

1.Google Audience Selection Dashboard

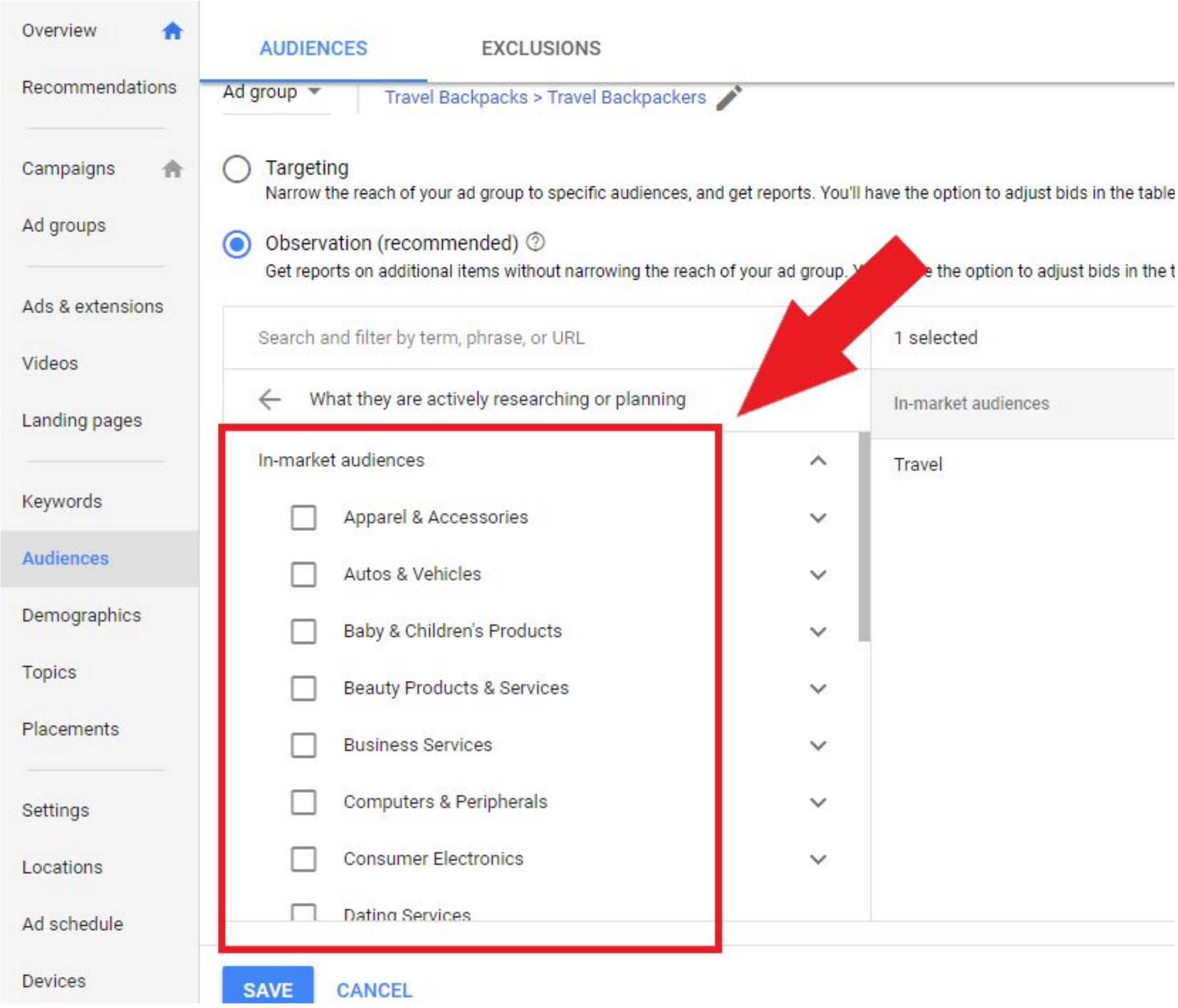

2. The summary page of text ad content (I collect content in red boxes)

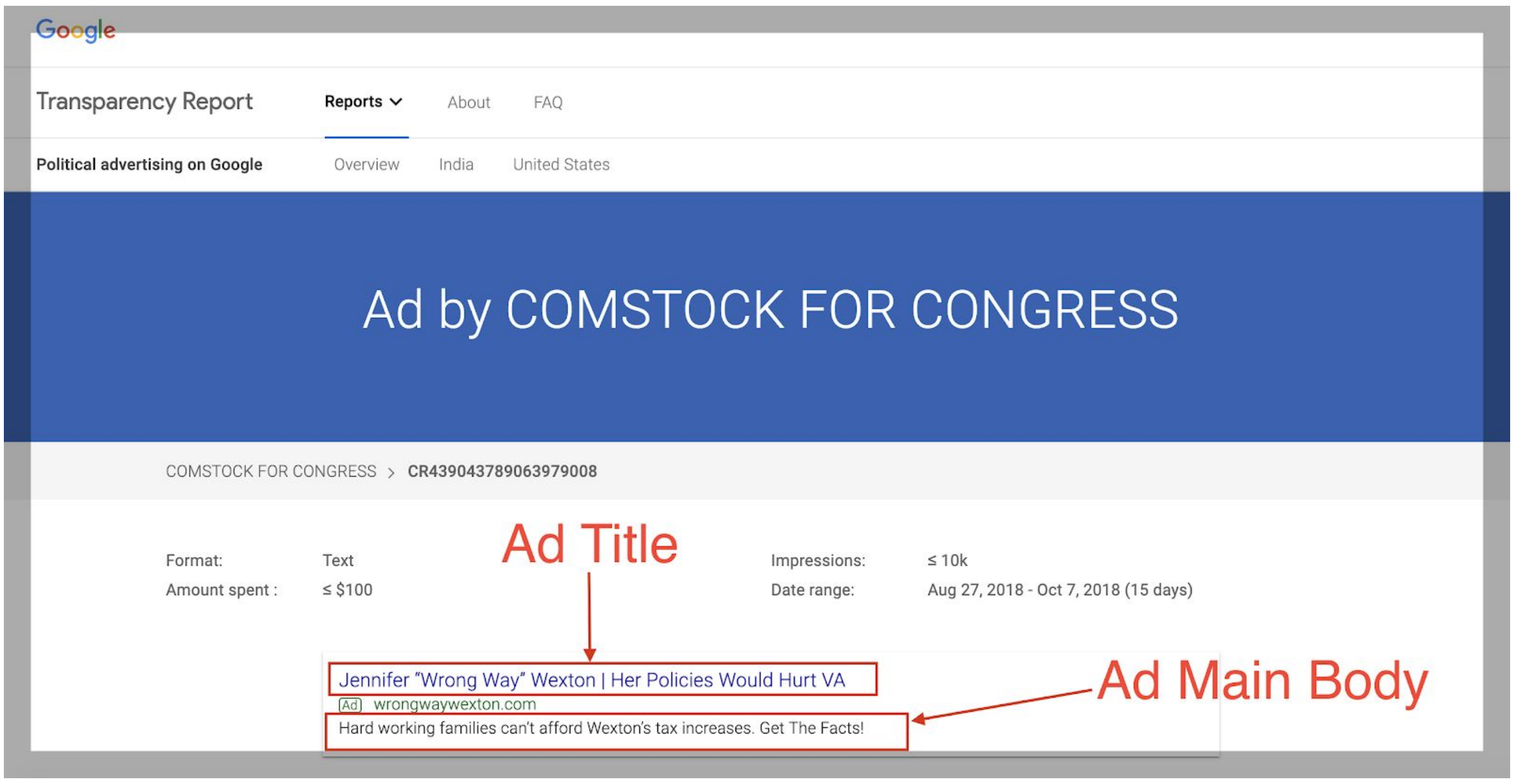

3. The summary page of image ad content (I download the image in red boxes):

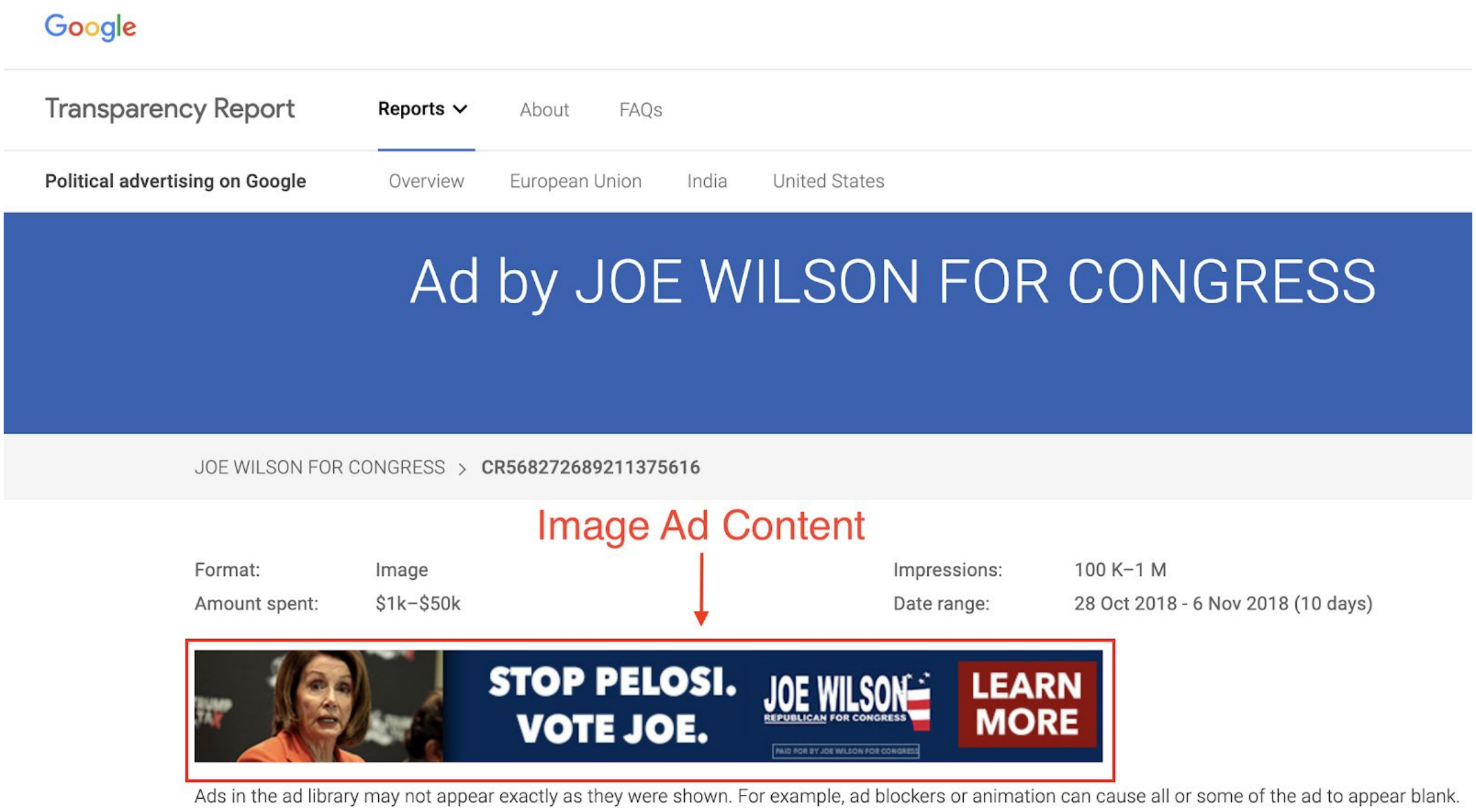

4. Summary Details of the NRC Emotion Lexicon:

| Association Lexicon | Version | # of Terms | Categories | Association Scores | Method of Creation | Papers |
|---|---|---|---|---|---|---|
| *Word–Emotion and Word–Sentiment Association Lexicon* | | | | | | |
| NRC Word–Emotion Association Lexicon (also called EmoLex)<br>README | 0.92<br>(2010) | 14,182 unigrams (words) | **sentiments:** negative, positive<br>**emotions:** anger, anticipation, disgust, fear, joy, sadness, surprise, trust | 0 (not associated) or 1 (associated) | **Manual:** By crowdsourcing on Mechanical Turk.<br>**Domain:** General | **Crowdsourcing a Word–Emotion Association Lexicon**, Saif Mohammad and Peter Turney, *Computational Intelligence,* 29 (3), 436–465, 2013.<br>Paper (pdf) BibTeX<br>**Emotions Evoked by Common Words and Phrases: Using Mechanical Turk to Create an Emotion Lexicon**, Saif Mohammad and Peter Turney, In *Proceedings of the NAACL-HLT 2010 Workshop on Computational Approaches to Analysis and Generation of Emotion in Text,* June 2010, LA, California.<br>Abstract Paper (pdf) Presentation |
| | | ~25,000 senses* | | not associated, weakly, moderately, or strongly associated | | |

5. Example of Image Recognition by Google Cloud Vision: detect labels

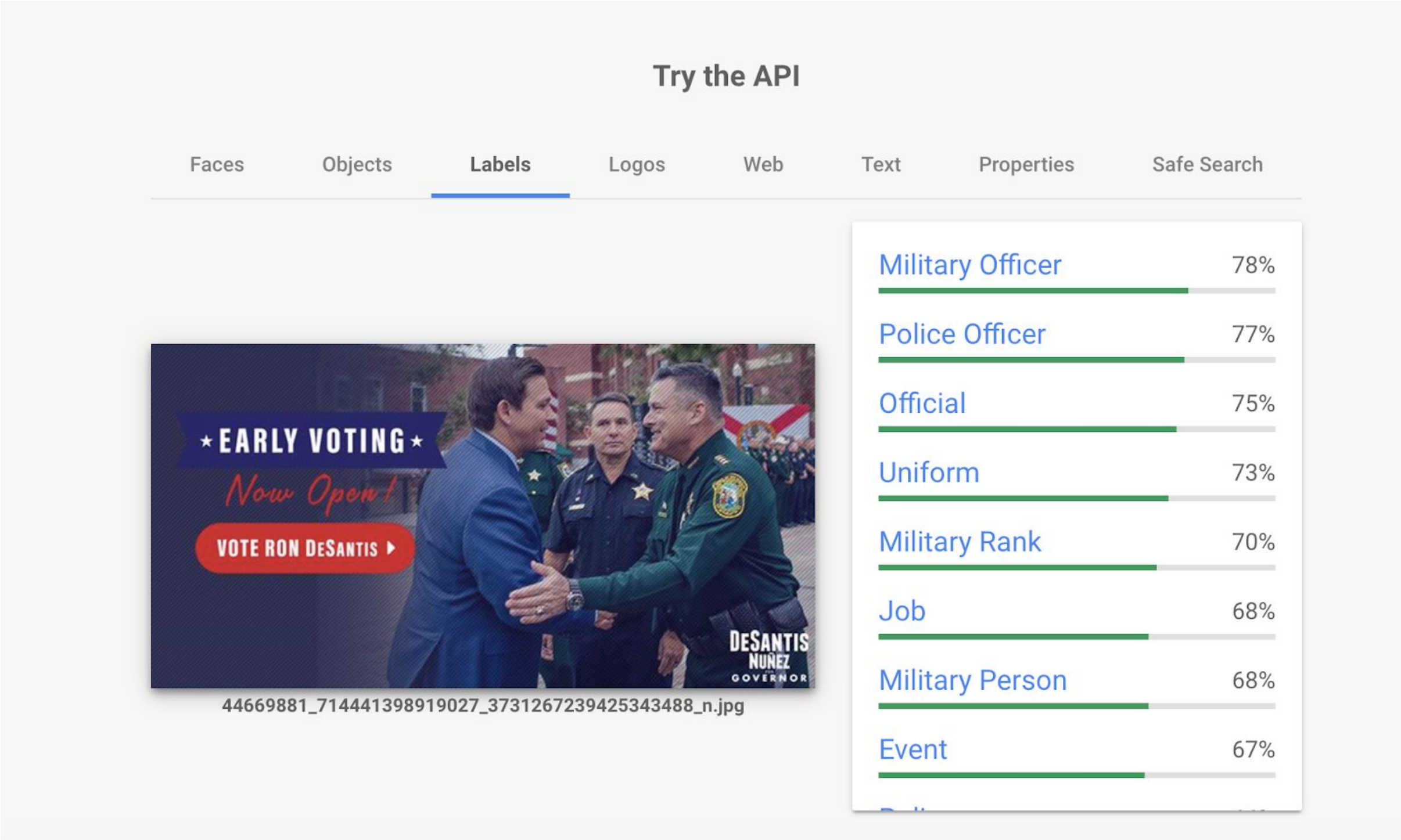

6. Example of Image Recognition by Google Cloud Vision: detect emotions

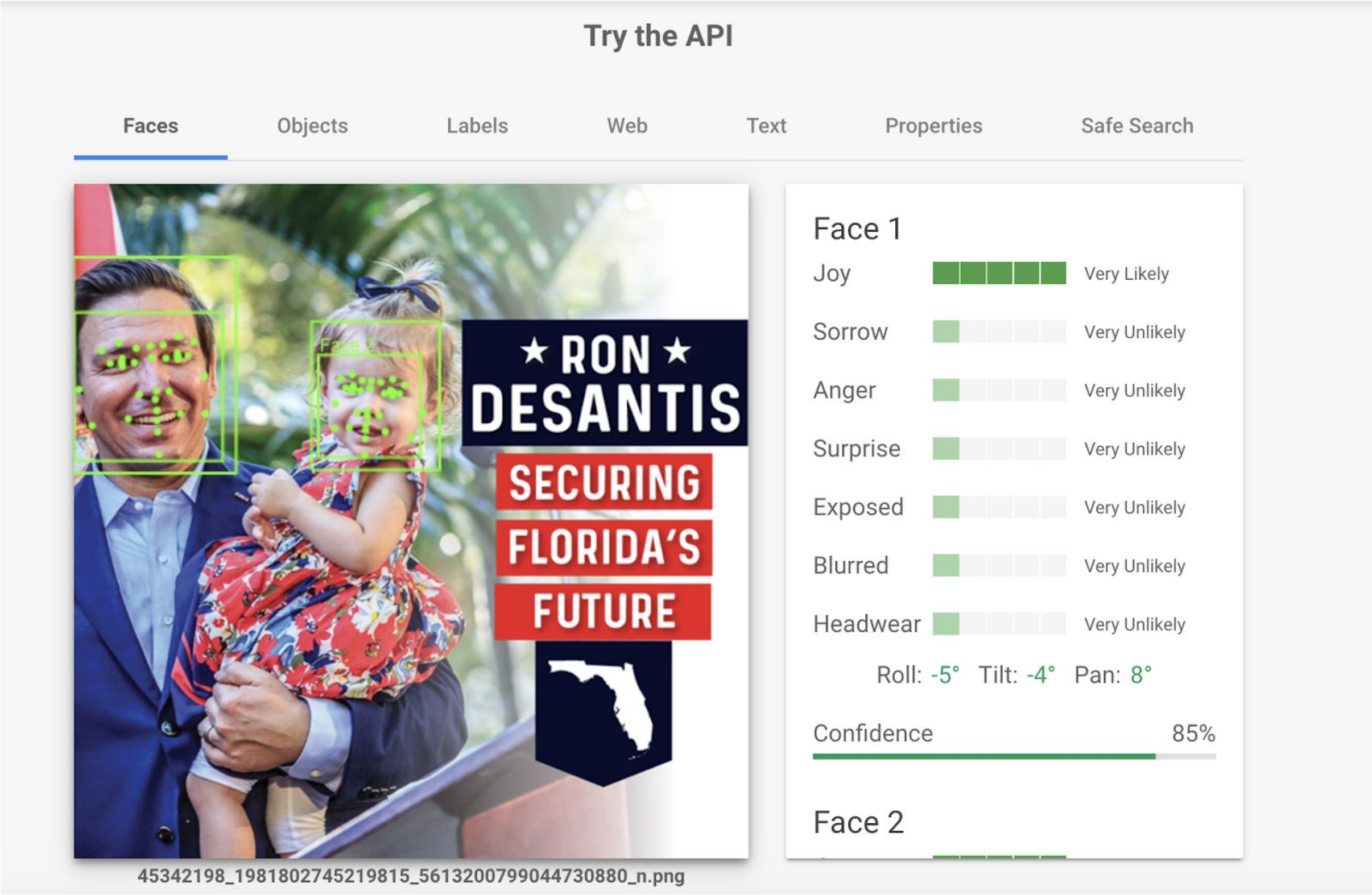

7. Two of the most frequent images that are detected as very likely to evoke “surprise”

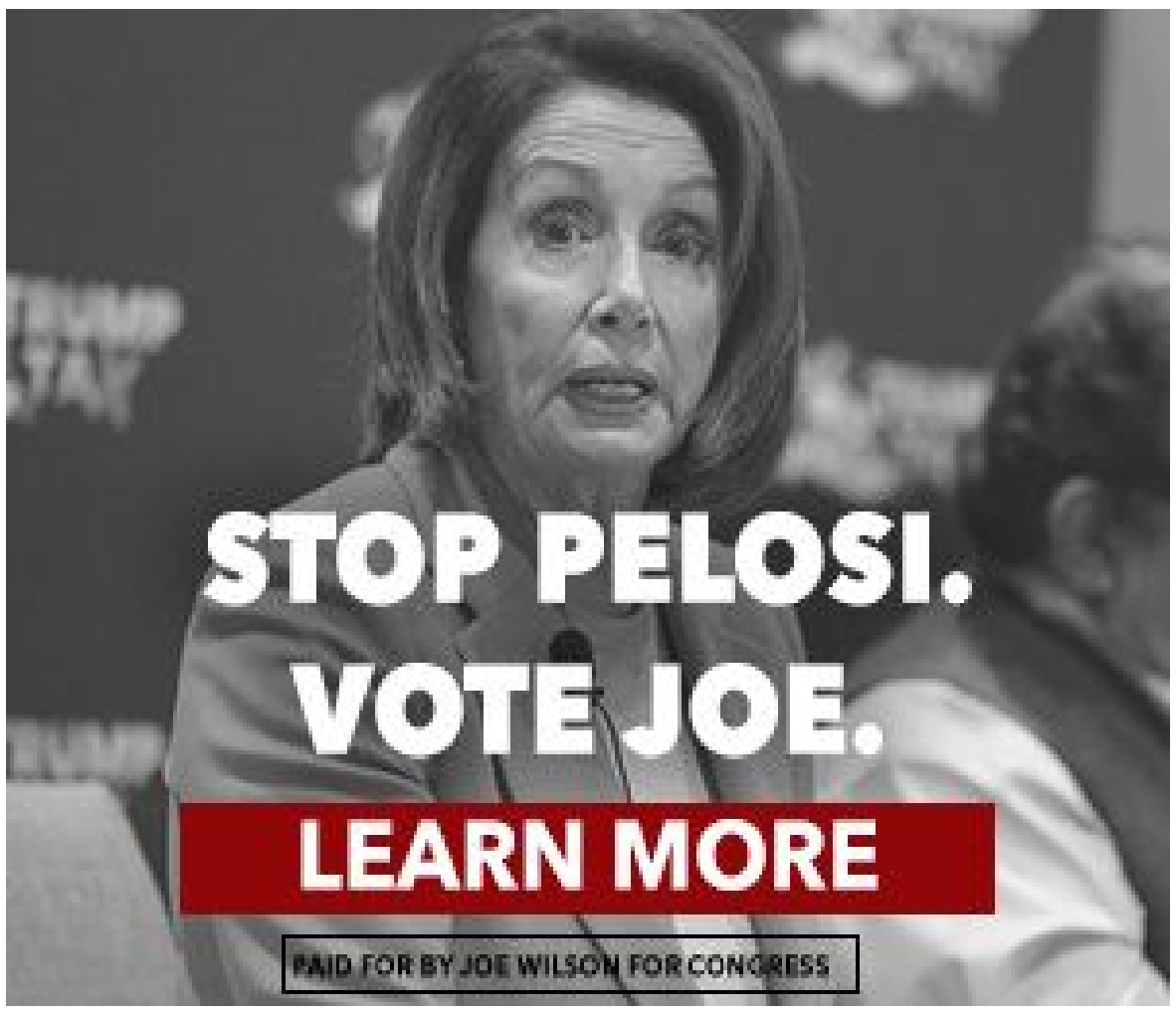

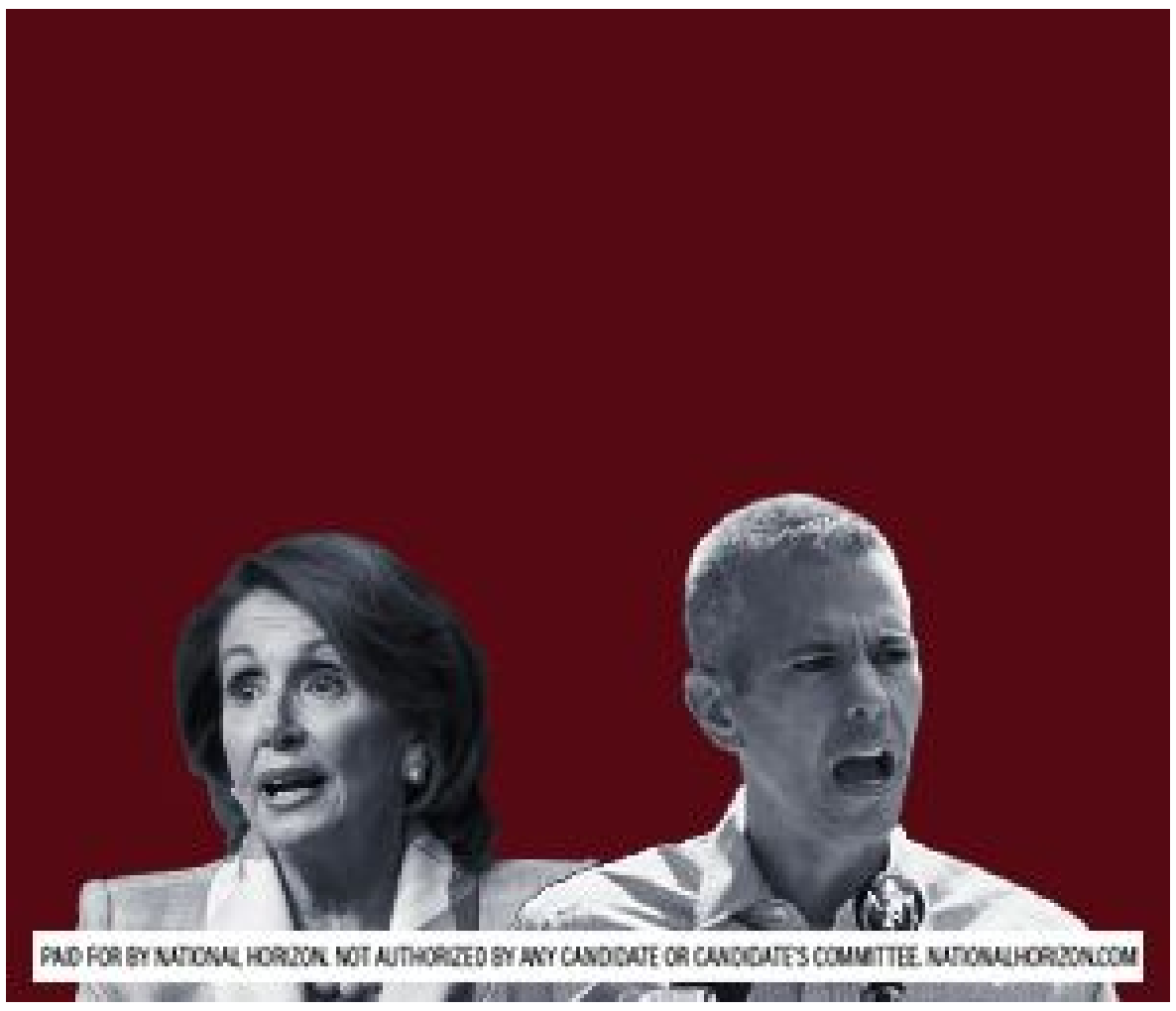

8. Topics by BTM and top terms in each topic

| Topic | Top Terms and probability |
|---|---|
| 1 | 1 Learn 0.03597306<br>2 Baldwin 0.03241249<br>3 Tammy 0.02686976<br>4 Stevens 0.02609892<br>5 families 0.02532808<br>6 Haley 0.02272189<br>7 UAW 0.02217129<br>8 tax 0.01769306<br>9 lose 0.01453627<br>10 much 0.01402237 |
| 2 | 1 Join 0.05992544<br>2 House 0.04350040<br>3 Taking 0.03893906<br>4 back 0.03351984<br>5 Left 0.03310137<br>6 one 0.03291305<br>7 Trump 0.03255735<br>8 thing 0.03247366<br>9 us 0.03236904<br>10 focused 0.03228535 |
| 3 | 1 campaign 0.03105182<br>2 today 0.03007659<br>3 money 0.02520042<br>4 Bredesen 0.02473231<br>5 Manchin 0.02223571<br>6 special 0.02157255<br>7 us 0.01911496<br>8 Phil 0.01817874<br>9 Help 0.01665738<br>10 Joe 0.01630629 |
| 4 | 1 Curbelo 0.03630148<br>2 Congress 0.03554422<br>3 Vote 0.03119000<br>4 Congressman 0.02740371<br>5 David 0.02276551<br>6 Carlos 0.02186627<br>7 Absentee 0.01803266<br>8 Iowa 0.01642348<br>9 Ballot 0.01339446<br>10 Voting 0.01187994 |
| 5 | 1 Learn 0.03268573<br>2 Congress 0.03171602<br>3 values 0.02874982<br>4 Golden 0.02749490<br>5 Jared 0.02726673<br>6 Matt 0.02715264<br>7 Rosendale 0.02658222<br>8 Heidi 0.02652518<br>9 Heitkamp 0.02224702<br>10 Montana's 0.02139138 |
| 6 | 1 Heller 0.06739790<br>2 Dean 0.05903799<br>3 Fund 0.03795326<br>4 Victory 0.03284853<br>5 Dirty 0.03262659<br>6 LCV 0.03262659<br>7 Learn 0.03225668<br>8 Marco 0.02974131<br>9 Record 0.02500649<br>10 Nevadans 0.02182529 |
| 7 | 1 Senate 0.08472073<br>2 Hawley 0.07256663<br>3 Josh 0.06334958<br>4 Trump 0.05033586<br>5 President 0.04009204<br>6 Donate 0.03732214<br>7 Corey 0.03701173<br>8 Support 0.03684458<br>9 Join 0.03538800 |
| | 10 Stewart 0.03350160 |
| 8 | 1 Mike 0.03042405<br>2 Bost 0.02745593<br>3 Illinois 0.02572453<br>4 Peter 0.01978829<br>5 Roskam 0.01871647<br>6 Congress 0.01822178<br>7 Learn 0.01813934<br>8 US 0.01698507<br>9 Tonette 0.01459408<br>10 small 0.01319247 |
| 9 | 1 John 0.06587959<br>2 Vote 0.05682461<br>3 Kingston 0.03676268<br>4 Senate 0.03209963<br>5 Steve 0.02261088<br>6 Congress 0.02212289<br>7 US 0.01924915<br>8 Primary 0.01691763<br>9 Republican 0.01686341<br>10 September 0.01545365 |
| 10 | 1 Beto 0.03447510<br>2 senator 0.02849295<br>3 Senate 0.02602608<br>4 Sinema 0.02454596<br>5 deserve 0.02288082<br>6 Dianne 0.02214076<br>7 de 0.02164739<br>8 today 0.01973556<br>9 campaign 0.01911885<br>10 Kyrsten 0.01874882 |
| 11 | 1 Scott 0.05838910<br>2 Rick 0.04005085<br>3 Donate 0.03689471<br>4 Senate 0.03662263<br>5 1st 0.02867787<br>6 Tarkanian 0.01703281<br>7 Vote 0.01605331<br>8 Oct 0.01589007<br>9 Danny 0.01583565<br>10 Team 0.01507382 |
| 12 | 1 Wexton 0.08458806<br>2 Wrong 0.06317737<br>3 Jennifer 0.04902769<br>4 Get 0.04688662<br>5 VA 0.04251139<br>6 Way 0.03363682<br>7 Facts 0.03059588<br>8 Direction 0.02839275 |
| | 9 Learn 0.02643786<br>10 Graham 0.02560005 |
| 13 | 1 Harris 0.08533870<br>2 Mark 0.07074740<br>3 Stop 0.04906889<br>4 healthcare 0.04444136<br>5 Vote 0.04227351<br>6 Marquez 0.03856315<br>7 Peterson 0.03531137<br>8 cut 0.03335197<br>9 Lea 0.02909965<br>10 coverage 0.02843261 |
| 14 | 1 Congress 0.03660709<br>2 care 0.02527530<br>3 health 0.02414212<br>4 fight 0.01478108<br>5 Diane 0.01428840<br>6 Learn 0.01300741<br>7 protect 0.01251473<br>8 Amy 0.01246546<br>9 McGrath 0.01103667<br>10 Delgado 0.01088886 |
| 15 | 1 Jon 0.03928849<br>2 Congress 0.03442279<br>3 Tester 0.03033352<br>4 Montana 0.02412199<br>5 us 0.01749635<br>6 Nunes 0.01646109<br>7 politics 0.01490821<br>8 Devin 0.01314827<br>9 choice 0.01294122<br>10 Lashar 0.01247535 |
| 16 | 1 Leah 0.06477357<br>2 Vukmir 0.05648815<br>3 WI 0.02799939<br>4 Senate 0.02677495<br>5 Congress 0.02012213<br>6 Gonzalez 0.01718346<br>7 US 0.01681612<br>8 Conservative 0.01604064<br>9 Learn 0.01477538<br>10 fight 0.01432642 |
| 17 | 1 Congress 0.03699664<br>2 Kim 0.02877999<br>3 Andy 0.02724735<br>4 Barbara 0.02592758<br>5 Us 0.02418207<br>6 Campaign 0.01937129<br>7 Davis 0.01800894<br>8 Paul 0.01720005<br>9 Spending 0.01583770<br>10 Vote 0.01507138 |
| 18 | 1 Congress 0.05761850<br>2 Dr 0.04647769<br>3 District 0.02904881<br>4 Ohio's 0.02904881<br>5 Healthcare 0.02863872<br>6 8th 0.02658827<br>7 MacArthur 0.02563139<br>8 Congressman 0.02460616<br>9 Enoch 0.02351258<br>10 Vanessa 0.02282910 |
| 19 | 1 Congress 0.05496623<br>2 Beto 0.02386343<br>3 Bryan 0.02200583<br>4 Steil 0.01967193<br>5 Kopser 0.01776670<br>6 Randy 0.01419441<br>7 Learn 0.01386099<br>8 Bryce 0.01362284<br>9 O'Rourke 0.01314653<br>10 Texas 0.01290838 |
| 20 | 1 McCaskill 0.04727213<br>2 Claire 0.04004434<br>3 Tina 0.03942775<br>4 Smith 0.03709842<br>5 Senate 0.03589949<br>6 now 0.03168614<br>7 Senator 0.02435559<br>8 seat 0.02387602<br>9 Give 0.01908033<br>10 Donate 0.01534654 |
| 21 | 1 Congress 0.09427658<br>2 Pramila 0.05815496<br>3 Learn 0.04738134<br>4 Mast 0.04518089<br>5 Jayapal 0.04400923<br>6 Brian 0.03272122<br>7 Fighting 0.03266406<br>8 protecting 0.02463386<br>9 Democrat 0.02143321 |
| | 10 Now 0.01903272 |
| 22 | 1 Donate 0.03816248<br>2 Scott 0.02923597<br>3 Steve 0.02588263<br>4 win 0.02271820<br>5 Daines 0.02106515<br>6 Democrat 0.01752288<br>7 Give 0.01676720<br>8 Senate 0.01539752<br>9 help 0.01511414<br>10 Walker 0.01383892 |
| 23 | 1 Diehl 0.12574657<br>2 Geoff 0.10526123<br>3 Senate 0.10291083<br>4 US 0.07097306<br>5 Warren 0.06226276<br>6 Elizabeth 0.05544199<br>7 Support 0.05518852<br>8 Help 0.05281507<br>9 Defeat 0.04864427<br>10 Massachusetts 0.04170829 |
| 24 | 1 Trump 0.10495479<br>2 President 0.05371737<br>3 Donald 0.04268662<br>4 Donate 0.03228864<br>5 Trump's 0.02864592<br>6 Official 0.02620035<br>7 Join 0.02006075<br>8 Support 0.01901753<br>9 Congress 0.01792301<br>10 Make 0.01534061 |
| 25 | 1 Vote 0.05309219<br>2 Congress 0.02929525<br>3 Election 0.02650527<br>4 vote 0.02564366<br>5 Day 0.02457690<br>6 Matt 0.02297676<br>7 Find 0.02108942<br>8 Nicholson 0.02051501<br>9 Kevin 0.01953031<br>10 28 0.01661723 |
| 26 | 1 Donate 0.05389975<br>2 Senate 0.05312574<br>3 McSally 0.04733236<br>4 Martha 0.04585470<br>5 Now 0.03037442<br>6 Today 0.02896712<br>7 US 0.02568343<br>8 Ted 0.02383048 |
| | 9 Support 0.02139117<br>10 Cruz 0.02136771 |
| 27 | 1 Susan 0.05843777<br>2 Learn 0.05219853<br>3 Hutchison 0.04038853<br>4 Maria 0.03894013<br>5 Cantwell 0.03821593<br>6 Senate 0.02880136<br>7 Dan 0.02635023<br>8 Congress 0.02629452<br>9 Feehan 0.02568174<br>10 served 0.02473471 |
| 28 | 1 Diehl 0.13931902<br>2 Geoff 0.09868897<br>3 Senate 0.08234259<br>4 US 0.07368373<br>5 Vote 0.06315156<br>6 Real 0.03996412<br>7 Elizabeth 0.03872912<br>8 Warren 0.03614812<br>9 candidate 0.03245700<br>10 Massachusetts 0.03068082 |
| 29 | 1 Jon 0.03792234<br>2 Tester 0.03496390<br>3 Learn 0.03286610<br>4 Roger 0.02146266<br>5 New 0.01656779<br>6 Trump 0.01613747<br>7 Get 0.01538442<br>8 First 0.01484652<br>9 veterans 0.01484652<br>10 Annie 0.01468515 |
| 30 | 1 Congress 0.07859721<br>2 Learn 0.07474708<br>3 Cisneros 0.04480166<br>4 Gil 0.04480166<br>5 District 0.04085431<br>6 Democrat 0.03708196<br>7 Conor 0.03511801<br>8 Lamb 0.03484578<br>9 Veteran 0.03443743<br>10 face 0.02924560 |

9. Links to repository that contain the codes and dataset required to replicate this research:

https://github.com/jam2399/Capstone_Project_An-analysis-of-Political-Ads-on-Google